\documentstyle[12pt,fleqn,epsf]{article}
\renewcommand{\theequation}{\arabic{equation}}

\begin{document}
\setlength{\unitlength}{1cm}
\setlength{\mathindent}{0cm}
\thispagestyle{empty}
\null
\hfill WUE-ITP-99-006\\
\null
\hfill UWThPh-1999-10\\
\null
\hfill HEPHY-PUB  709/99\\
\null
\hfill hep-ph/9903220\\
\vskip .8cm
\begin{center}
{\Large \bf Polarization and Spin Effects in \\[.5em]
Neutralino Production and Decay%
}
\vskip 2.5em
{\large
{\sc G.~Moortgat-Pick$^{a}$\footnote{e-mail:
    gudi@physik.uni-wuerzburg.de},  
H. Fraas$^{a}$\footnote{e-mail:
    fraas@physik.uni-wuerzburg.de},
A.~Bartl$^{b}$\footnote{e-mail:
     bartl@merlin.ap.univie.ac.at},
W.~Majerotto$^{c}$\footnote{e-mail:
     majer@qhepu3.oeaw.ac.at.}%
}}\\[1ex]
{\normalsize \it
$^{a}$ Institut f\"ur Theoretische Physik, Universit\"at
W\"urzburg, Am Hubland, D-97074~W\"urzburg, Germany}\\
{\normalsize \it
$^{b}$ Institut f\"ur Theoretische Physik, Universit\"at Wien, 
Boltzmanngasse 5, A-1090 Wien}\\
{\normalsize \it
$^{c}$ Institut f\"ur Hochenergiephysik, \"Osterreichische
 Akademie der Wissenschaften, Nikolsdorfergasse 18 
, A-1050 Wien}
\vskip 1em
\end{center} \par
\vskip .8cm

\begin{abstract}
We study
 the production of neutralinos 
$e^+ e^- \to\tilde{\chi}^{0}_i\tilde{\chi}^{0}_j$ with
 polarized beams and the subsequent  decays $\tilde{\chi}^{0}_{i} \to 
\tilde{\chi}^0_k\ell^{+}\ell^{-}$ and 
$\tilde{\chi}^{0}_{j} \to \tilde{\chi}^0_l \ell^{+}\ell^{-}$, 
including the complete spin 
correlations between production and decay.
We present analytical formulae for the differential
 cross section of the combined process of production and decay of neutralinos.
We also allow for complex couplings.
The spin correlations have a strong influence on the decay angular
distributions and the corresponding forward--backward asymmetries. 
They are very sensitive to the SUSY
parameters and depend strongly on the beam polarizations. We present
numerical results for the cross section and the electron
forward--backward asymmetry for 
$e^+ e^- \to\tilde{\chi}^{0}_1\tilde{\chi}^{0}_2$,
 $\tilde{\chi}^{0}_{2} \to 
\tilde{\chi}^0_1 e^{+} e^{-}$. We study the dependence on the
parameter $M'$ for various mass splittings between $\tilde{e}_L$ and
$\tilde{e}_R$ and different beam polarizations.
\end{abstract}
\newpage
\section{Introduction}
\vspace{-.3cm}
The search for supersymmetric (SUSY) particles is one of the main goals of
present and future colliders. 
In particular, an $e^{+}e^{-}$ linear collider  
will be an excellent discovery machine for SUSY particles
\cite{JLC}. 
Experiments at a linear collider will also allow us to 
determine precisely the parameters of the underlying SUSY
model \cite{accomando}.

The neutralinos, the
supersymmetric partners of the neutral gauge and Higgs bosons, are of
particular interest as they are expected to be 
relatively light. Most studies of neu\-tralino production 
$e^{+}e^{-}\to\tilde{\chi}^0_i\tilde{\chi}^0_j$, i, j=1,$\ldots$,4, and
decays have been performed in the Minimal Supersymmetric
Standard Model (MSSM) (see \cite{bartl, ambrosanio, desy}, and references
therein).
For a clear identification of neutralinos a
precise analysis of their decay characteristics is
indispensable. 
By measuring cross sections, branching ratios, various energy and 
angular distributions of the decay products of
the neutralinos, one obtains valuable information about the 
parameters of the MSSM.
 
Since decay angular distributions depend on the polarization of the parent
particles one has to take into account the spin correlations between
production and decay of the neutralinos. In general, quantum mechanical
interference effects between various polarization states of the
decaying particles preclude simple factorization of the differential
cross section into a production and a decay factor \cite{tsai,tata}, 
unless the production amplitude is dominated by a single spin
component \cite{feng}. A variety of event generators for production
and decay of SUSY particles, such as DFGT, 
SUSYGEN, GRACE and CompHEP \cite{dfgt},  have been developed which 
include spin correlations between production and decay.

In a previous paper \cite{prd97} the process 
$e^{+}e^{-}\to\tilde{\chi}^0_i\tilde{\chi}^0_j$, i, j=1,$\ldots$,4,  
with unpolarized beams and the subsequent leptonic decays
$\tilde{\chi}^0_i\to\tilde{\chi}^0_k\ell^{+}\ell^{-}$,
$\tilde{\chi}^0_j\to\tilde{\chi}^0_l \ell^{+} \ell^{-}$ have been
studied with complete spin correlations.
Some results for polarized beams have been presented in
\cite{cracow98}. In the present paper we give the complete analytical
formulae for polarized beams. We fully include
the spin correlations between production and decay. The formulae are
given in a transparent form in the laboratory system 
(which is identical to the overall CMS)
 in terms of the basic kinematic
variables. Moreover, we include complex couplings allowing for studies 
of CP violating phenomena.
The expression for the differential cross section is composed of the joint spin
density matrix for the production of neutralinos 
and the decay matrices for their
leptonic decays.  Our formulae can easily be extended to hadronic
decays.

The masses and couplings of the neutralinos depend on the MSSM
parameter $M'$, $M$, $\mu$ and $\tan\beta$. The parameters 
$M$, $\mu$ and $\tan\beta$ 
can in principle be determined by chargino production alone
\cite{choi,epj98}. The cross section for chargino production
with polarized beams and the decay angular distributions also
give information on the sneutrino mass $m_{\tilde{\nu}}$
\cite{cracow99}. However, a precise
determination of the parameter $M'$ is only possible in the neutralino 
sector. A study of neutralino production and decay also gives
information about the masses of the left and right selectrons, 
$m_{\tilde{e}_L}$ and $m_{\tilde{e}_R}$.

It is known that the forward--backward asymmetry of the production 
process $e^{+}e^{-}\to \tilde{\chi}^0_i \tilde{\chi}^0_j$ vanishes due 
to the Majorana nature of the neutralinos \cite{bartl,petcov,christova}.
However, taking into account neutralino decay, for instance
$\tilde{\chi}^0_i\to \ell^{+} \ell^{-} \tilde{\chi}^0_k$, the
forward--backward asymmetry $A_{FB}$ of one of the decay leptons does not 
  vanish \cite{prd97, cracow98}. 
This is a consequence of spin correlations between
  production and decay. As we shall show 
  this $A_{FB}$ depends very sensitively on
  the SUSY parameters. Furthermore, it depends very
  strongly on the beam polarization.
The forward--backward asymmetry $A_{FB}$ of the decay lepton is due to
a complex interplay of the  $Z$ and $\tilde{\ell}_{L}$,
$\tilde{\ell}_R$ exchange
amplitudes in production and decay, where the polarization  of
$\tilde{\chi}^0_i$ plays a crucial r\^ole. The polarization vector of 
$\tilde{\chi}^0_i$ is determined by the characteristics of the
production process and strongly influences the decay distribution.

The main purpose of our paper is the presentation of the formulae for
the combined process $e^{+}e^{-}\to \tilde{\chi}^0_i
\tilde{\chi}^0_j$, $\tilde{\chi}^0_i\to \ell^{+}\ell^{-}
\tilde{\chi}^0_k$ and $\tilde{\chi}^0_j\to \ell^{+}\ell^{-}
\tilde{\chi}^0_l$, with both beams polarized. We also present
numerical results for the cross section and the lepton forward--backward
asymmetry $A_{FB}$ of
$e^{+}e^{-}\to\tilde{\chi}^0_2\tilde{\chi}^0_1$,
$\tilde{\chi}^0_2\to\ell^{+}\ell^{-}\tilde{\chi}^0_1$. In particular,
we study their dependence on $M'$,
$m_{\tilde{\ell}_L}$ and $m_{\tilde{\ell}_R}$, and on the beam
polarization.

In Section~2 we present the formalism used. In Section~3 we
give the formulae for the spin density production matrix of the
neutralinos in the laboratory system for polarized beams. 
In Section~4 we give the
decay matrices for the leptonic decay of $\tilde{\chi}^0_i$ and 
$\tilde{\chi}^0_j$ in covariant form.
In the Section~5 we present our numerical results for the cross section
and the forward--backward asymmetry $A_{FB}$
of the decay lepton as a function of
the parameter $M'$ for various slepton masses 
$m_{\tilde{\ell}_L},m_{\tilde{\ell}_R}$ and for different beam 
polarizations.
\section{Definitions and formalism}
We give the analytical formulae for the differential cross section 
of neutralino production
\begin{equation}
e^{-}(p_1)+e^{+}(p_2)\to \tilde{\chi}^0_i(p_3)+\tilde{\chi}^0_j(p_4),
\label{eq_1}
\end{equation}
with polarized beams and 
the subsequent leptonic decays 
\begin{eqnarray}
\tilde{\chi}^0_i(p_3)&\to&
\tilde{\chi}^0_k(p_5)+\ell^{+}(p_6)+\ell^{-}(p_7),\label{eq_2}\\
\tilde{\chi}^0_j(p_4)&\to&
\tilde{\chi}^0_l(p_8)+\ell^{+}(p_9)+\ell^{-}(p_{10})\label{eq_3},
\end{eqnarray}
with complete spin correlations between production and decay.
\subsection{Lagrangian and couplings}
The parts of the interaction Lagrangian of the MSSM  relevant for our
study are (in our notation and
conventions we follow closely \cite{2haber}):
\begin{eqnarray}
& & {\cal L}_{Z^0 \ell^{+} \ell^{-}} =
-\frac{g}{\cos\theta_W}Z_{\mu}\bar{\ell}\gamma^{\mu}[L_{\ell}P_L+
 R_{\ell}P_R]\ell, \\
& & {\cal L}_{Z^0\tilde{\chi}^0_i\tilde{\chi}^0_j} =
\frac{1}{2}\frac{g}{\cos\theta_W}Z_{\mu}\bar{\tilde{\chi}}^0_i\gamma^{\mu}
[O_{ij}^{''L} P_L+O_{ij}^{''R} P_R]\tilde{\chi}^0_j,\\
& & {\cal L}_{\ell \tilde{\ell}\tilde{\chi}^0_i} =
g f^L_{\ell i}\bar{\ell}P_R\tilde{\chi}^0_i\tilde{\ell}_L+
g f^R_{\ell i}\bar{\ell}P_L\tilde{\chi}^0_i\tilde{\ell}_R+\mbox{h.c.},
 \quad i, j=1,\cdots,4.\label{eq_4}
\end{eqnarray}

The couplings are:
\begin{eqnarray}
L_{\ell}&=&T_{3\ell}-e_{\ell}\sin^2\theta_W, \quad
 R_{\ell}=-e_{\ell}\sin^2\theta_W \label{eq_5},\\
f_{\ell i}^L &=& -\sqrt{2}\bigg[\frac{1}{\cos
\theta_W}(T_{3\ell}-e_{\ell}\sin^2\theta_W)N_{i2}+e_{\ell}\sin \theta_W
N_{i1}\bigg],\nonumber\\
f_{\ell i}^R &=& -\sqrt{2}e_{\ell} \sin \theta_W\Big[\tan \theta_W N_{i2}^*-
                                                           N_{i1}^*\Big],
\label{eq_6}\\
O_{ij}^{''L}&=&-\frac{1}{2}\Big(N_{i3}N_{j3}^*-N_{i4}N_{j4}^*\Big)\cos2\beta
  -\frac{1}{2}\Big(N_{i3}N_{j4}^*+N_{i4}N_{j3}^*\Big)\sin2\beta,\nonumber\\
O_{ij}^{''R}&=&-O_{ij}^{''L*},\mbox{ with } i\mbox{, }j=1,\ldots,4.\label{eq_7}
\end{eqnarray}
where $P_{L, R}=\frac{1}{2}(1\mp \gamma_5)$, $g$ is the weak coupling
constant ($g=e/\sin\theta_W$, $e>0$), and $e_\ell$ and $T_{3 \ell}$ denote the
charge and the third component of the weak isospin of the lepton
$\ell$, $\tan \beta=v_2/v_1$ is the ratio of
 the vacuum expectation values of the two neutral Higgs fields.
$N_{ij}$ is the unitary $4\times 4$ matrix which diagonalizes
the neutral gaugino-higgsino mass matrix $Y_{\alpha\beta}$, 
 $N_{i \alpha }Y_{\alpha\beta}N_{k \beta}=m_{\tilde{\chi}^0_i}\delta_{ik}$.
We use the basis $\tilde{\gamma},
\tilde{Z}, \tilde{H}^0_a, \tilde{H}^0_b$ \cite{bartl}.
\subsection{CP conserving and CP violating case}
In the formulae for the cross section we shall present in the
following, one has to distinguish between two cases, CP conservation
and CP violation:
If CP is conserved the neutralino mass matrix $Y_{\alpha\beta}$ 
is real and the matrix
$N_{ij}$ can be chosen real and orthogonal. Then all the
couplings given in eqs.~(\ref{eq_6}), (\ref{eq_7}) are real. In this
case some of the mass eigenvalues may be negative. We therefore write
the eigenvalues in the form 
$m_{\tilde{\chi}^0_i}=\eta_i m_i$, $i=1,\ldots,4$, with $m_i\ge0$
and $\eta_i=\pm 1$ \cite{bartl}. $\eta_i$ is related to the CP
eigenvalue of the neutralino $\tilde{\chi}^0_i$ \cite{ellis}.

If CP is violated the neutralino mass matrix is complex and the
matrix $N_{ij}$ is complex and unitary. In this case the
diagonalization of the mass matrix is done with the singular value
decomposition method, $N_{i \alpha}Y_{\alpha \beta} N_{k \beta}=m_i
\delta_{ik}$, $m_i\ge 0$.
In this method all masses $m_i$ 
are chosen positive. The neutralino couplings, 
eqs.~(\ref{eq_6}), (\ref{eq_7}), are complex.

The formulae given below are applicable to both cases. In the case of
CP conservation the imaginary parts of all couplings are zero and the
sign $\eta_i$ of the mass eigenvalues, appearing explicitly in the
formulae, has to be taken into account.

In the case of CP violation the imaginary parts of the couplings do
not vanish. All factors $\eta_i$ appearing in the formulae have to
be set $\eta_i=+1$.
\subsection{Helicity amplitudes and cross section}
For the calculation of the amplitude of the combined processes of 
neutralino production and decays, eqs.~(\ref{eq_1})--(\ref{eq_3}),  
we use the same formalism as for
the chargino production and decays \cite{epj98},
following the method of \cite{haber}.
The helicity amplitudes for the production eq.~(\ref{eq_1}) are
\begin{equation}
T_P^{\lambda_i\lambda_j}=T_P^{\lambda_i\lambda_j}(s,Z)
+T_P^{\lambda_i\lambda_j}(t,\tilde{e}_L)
+T_P^{\lambda_i\lambda_j}(t,\tilde{e}_R)
+T_P^{\lambda_i\lambda_j}(u,\tilde{e}_L)
+T_P^{\lambda_i\lambda_j}(u,\tilde{e}_R),\label{eq_8}
\end{equation}
and those for the decays, eq.(\ref{eq_2}) and eq.(\ref{eq_3}) are
\begin{eqnarray}
T_{D,\lambda_i}&=&T_{D,\lambda_i}(s_i,Z)+T_{D,\lambda_i}(t_i,\tilde{\ell}_L)
+T_{D,\lambda_i}(t_i,\tilde{\ell}_R)+T_{D,\lambda_i}(u_i,\tilde{\ell}_L)
+T_{D,\lambda_i}(u_i,\tilde{\ell}_R),\label{eq_9}\\
T_{D,\lambda_j}&=&T_{D,\lambda_j}(s_j,Z)+T_{D,\lambda_j}(t_j,\tilde{\ell}_L)
+T_{D,\lambda_j}(t_j,\tilde{\ell}_R)+T_{D,\lambda_j}(u_j,\tilde{\ell}_L)
+T_{D,\lambda_j}(u_j,\tilde{\ell}_R).\label{eq_10}
\end{eqnarray}
They correspond to the Feynman diagrams in
Fig.~1, and are given in the Appendix~A, eqs.~(\ref{eq_a1})--(\ref{eq_a7}).

We introduce the kinematic variables
$s=(p_1+p_2)^2$, $t=(p_1-p_4)^2$, and $u=(p_1-p_3)^2$ for the production
process, eq.~(\ref{eq_1}), and 
$s_i=(p_6+p_7)^2$, $t_i=(p_3-p_6)^2$, $u_i=(p_3-p_7)^2$ for the decay
process of the neutralino $\tilde{\chi}^0_i$, eq.~(\ref{eq_2}), 
and $s_j=(p_9+p_{10})^2$, $t_j=(p_4-p_9)^2$ and $u_j=(p_4-p_{10})^2$
for the decay of the neutralino $\tilde{\chi}^0_j$, eq.~(\ref{eq_3}), 
with the particle
momenta $p_k$ as denoted in eqs.~(\ref{eq_1})--({\ref{eq_3}).

The amplitude for the whole process is
\begin{equation}
T=\Delta(\tilde{\chi}^{0}_i) \Delta(\tilde{\chi}^{0}_j) 
\sum_{\lambda_i, \lambda
_j}T_P^{\lambda_i \lambda_j}T_{D, \lambda_i} T_{D, \lambda_j}\label{eq_12}, 
\end{equation}
where $\Delta(\tilde{\chi}^{0}_{i})=1/[p^2_{3}-m_{i}^2
+im_{i}\Gamma_{i}]$, $m_{i}$, $\Gamma_{i}$,  
and
$\Delta(\tilde{\chi}^{0}_{j})=1/[p^2_{4}-m_{j}^2
+im_{j}\Gamma_{j}]$, $m_{j}$, $\Gamma_{j}$ 
denote the propagator, mass and width of
$\tilde{\chi}^{0}_{i}$ and $\tilde{\chi}^{0}_{j}$, respectively. 
For these propagators we use the narrow width approximation.

\noindent The differential cross section in the laboratory system 
is then given by:
\begin{equation}
d\sigma=\frac{1}{8 E_b^2}|T|^2 (2\pi)^4
\delta^4(p_1+p_2-\sum_{i} p_i) d{\rm lips}(p_3\ldots p_{10})\label{eq_13},
\end{equation}
$E_b=\sqrt{s}/2$ denotes the beam energy and $d{\rm lips}(p_3,\ldots,p_{10})$ is
the Lorentz invariant phase space element.

The amplitude squared $|T|^2$ of the combined processes of production 
and decays, eq.~(\ref{eq_12}),  is given by: 
\begin{eqnarray}
|T|^2& = &4 |\Delta(\tilde{\chi}^{0}_i)|^2|\Delta(\tilde{\chi}^{0}_j)|^2
         \Big(P D(\tilde{\chi}^0_i) D(\tilde{\chi}^0_j)
    +\sum^3_{a=1}\Sigma_P^a(\tilde{\chi}^0_i) \Sigma_D^a(\tilde{\chi}^0_i) 
D(\tilde{\chi}^0_j)\nonumber\\
& &\phantom{4|\Delta(\tilde{\chi}^{0}_i)|^2|\Delta(\tilde{\chi}^{0}}
+\sum^3_{b=1}\Sigma_P^b(\tilde{\chi}^0_j) \Sigma_D^b(\tilde{\chi}^0_j)
D(\tilde{\chi}^0_i)
    +\sum^3_{a,b=1}\Sigma_P^{ab}(\tilde{\chi}^0_i\tilde{\chi}^0_j)
 \Sigma^a_D(\tilde{\chi}^0_i) 
\Sigma^b_D(\tilde{\chi}^0_j)\Big).
\label{eq_14}
\end{eqnarray}
Here a, b=1, 2, 3 refer to the polarization vectors of
$\tilde{\chi}^0_i$ and $\tilde{\chi}^0_j$ defined in 
eqs.~(\ref{eq_21})--(\ref{eq_26}) below. 
If one neglects all spin correlations between 
production and decay
only the first term in eq.~(\ref{eq_14}) 
would contribute. 
The second and third term describe the spin
correlations between
the production and the decay processes 
$\tilde{\chi}_i^0 \to \tilde{\chi}^0_k \ell^+ \ell^-$ and 
$\tilde{\chi}^0_j \to \tilde{\chi}^0_\ell \ell^+ \ell^-$, respectively,
because $\Sigma^a_P(\tilde{\chi}^0_i) (\Sigma^b_P(\tilde{\chi}^0_j))$
as well as $\Sigma^a_D(\tilde{\chi}^0_i) (\Sigma^b_D(\tilde{\chi}^0_j))$
depend on the polarization of the neutralino $\tilde{\chi}^0_i 
(\tilde{\chi}^0_j)$. 
Since $\Sigma^{ab}_P(\tilde{\chi}^0_i \tilde{\chi}^0_j)$ depends on the
polarizations of both neutralinos 
the last term is due to spin-spin correlations between both decaying
neutralinos (see Appendix eq.~(\ref{B_5})).

Owing to the Majorana character the spin correlations do not influence 
the energy distribution of the neutralino decay
products and the opening angle distribution between the leptons from
the decay of one of the neutralinos. Therefore, these distributions are 
given only by the first term 
$P D(\tilde{\chi}^0_i)D(\tilde{\chi}^0_j)$ in eq.~(\ref{eq_14}) 
\cite{prd97,cracow98}.
 
\noindent We give the explicit expressions for
$P$, $\Sigma_P^a(\tilde{\chi}^0_i)$, $\Sigma_P^b(\tilde{\chi}^0_j)$,
$\Sigma_P^{ab}(\tilde{\chi}^0_i\tilde{\chi}^0_j)$ in Section~3
and for the quantities
$D(\tilde{\chi}^0_i)$, $\Sigma_D^a(\tilde{\chi}^0_i)$,
$D(\tilde{\chi}^0_j)$, $\Sigma_D^b(\tilde{\chi}^0_j)$ 
in Section~4.

\section{Spin-density production matrix}
\begin{sloppypar}
In this section we give the analytical formulae for the
quantities $P$, $\Sigma_P^a(\tilde{\chi}^0_i)$, 
$\Sigma^b_P(\tilde{\chi}^0_j)$, 
$\Sigma_P^{ab}(\tilde{\chi}^0_i,\tilde{\chi}^0_j)$, eq.~(\ref{eq_14}),
for the production in the laboratory system. 
\end{sloppypar}

It is useful
to introduce the abbrevations  
$c_{L}(\alpha\beta), c_{R}(\alpha\beta)$ as shown in Table~\ref{tab_1}.
$P_{-}^3 (P_{+}^3)$
is the longitudinal beam polarization of $e^{-} (e^{+})$,
and $L_{\ell} (R_{\ell})$ is defined in eq.~(\ref{eq_5}).
The arguments $\alpha$, $\beta$
denote the exchanged particles. 
Generally $c_L(\alpha\beta)$ $(c_R(\alpha\beta))$
is large for $P^3_{-}<0$, $P^3_{+}>0$ ($P^3_{-}>0$, $P^3_{+}<0$), 
and favours left(right) selectron exchange. 

\begin{table}
\begin{tabular}{|l||c|c|}
\hline
$(\alpha\beta)$ & $c_L(\alpha\beta)$ & $c_R(\alpha\beta)$\\\hline\hline
$(Z^0 Z^0)$ & $L_{\ell}^2 (1-P^3_{-})(1+P_{+}^3)$ 
& $R_{\ell}^2 (1+P^3_{-})(1-P_{+}^3)$\\\hline
$(Z^0 \tilde{e}_L)$ & $L_{\ell} (1-P^3_{-})(1+P_{+}^3)$ & 0\\\hline
$(Z^0 \tilde{e}_R)$ & 0 & $R_{\ell} (1+P^3_{-})(1-P_{+}^3)$\\\hline
$(\tilde{e}_L \tilde{e}_L)$ & $(1-P^3_{-})(1+P_{+}^3)$ & 0\\\hline
$(\tilde{e}_R \tilde{e}_R)$ & 0 & $(1+P^3_{-})(1-P_{+}^3)$\\\hline
\end{tabular}
\caption{$c_L(\alpha\beta), c_R(\alpha\beta)$ for longitudinal
beam polarization $P^3_{-} (P_{+}^3)$ of $e^{-} (e^{+})$,
$\alpha$, $\beta$ denote the exchanged particles,
$L_{\ell}$, $R_{\ell}$ are defined in eq.~(\ref{eq_5}). 
For unpolarized beams one has 
$P^3_{-}=P^3_{+}=0$.\label{tab_1}}
\end{table}
\subsection{Neutralino polarization independent quantities}
The expression $P$ of eq.~(\ref{eq_14}) is independent of the 
neutralino polarization and reads: 
\begin{equation}
P=P(Z Z)+P(Z \tilde{e}_L)+P(Z \tilde{e}_R)+
P(\tilde{e}_L \tilde{e}_L)+P(\tilde{e}_R \tilde{e}_R),\label{eq_15}
\end{equation}
with
\begin{eqnarray}
P(Z Z)&=&
2 \frac{g^4}{\cos^4\Theta_W}|\Delta^s(Z)|^2
[c_{R}(ZZ)+c_{L}(ZZ)]   E_b^2 
\nonumber\\& &
\Big\{ |O^{''L}_{ij}|^2 (E_i E_j+q^2\cos^2\Theta)-[(Re O^{''L}_{ij})^2 
-(Im O^{''L}_{ij})^2] \eta_i \eta_j
m_i m_j\Big\}\label{eq_16},\\
P(Z \tilde{e}_L)&=&
\frac{g^4}{\cos^2 \Theta_W} c_L(Z \tilde{e}_L) E_b^2
Re\Big\{\Delta^s(Z)
\nonumber\\& & 
\Big[
-(\Delta^{t*}(\tilde{e}_L) f^{L*}_{\ell i} 
f^{L}_{\ell j} O^{''L*}_{ij}
+\Delta^{u*}(\tilde{e}_L) f^L_{\ell i} f^{L*}_{\ell j}
O^{''L}_{ij}) \eta_i \eta_j m_i m_j
\nonumber\\& & 
-(\Delta^{t*}(\tilde{e}_L) f^{L*}_{\ell i} f^{L}_{\ell j} 
O^{''L}_{ij}
-\Delta^{u*}(\tilde{e}_L) f^L_{\ell i} f^{L*}_{\ell j} 
O^{''L*}_{ij})
2 E_b q \cos\Theta
\nonumber\\
& &
+(\Delta^{t*}(\tilde{e}_L) f^{L*}_{\ell i} f^{L}_{\ell j}
 O^{''L}_{ij}
+\Delta^{u*}(\tilde{e}_L) f^L_{\ell i} f^{L*}_{\ell j} 
O^{''L*}_{ij})
(E_i E_j+q^2\cos^2\Theta)
\Big]\Big\},\label{eq_17}\\
P(\tilde{e}_L \tilde{e}_L)&=&
\frac{g^4}{4} c_L(\tilde{e}_L\tilde{e}_L) E_b^2 
\Big\{|f^L_{\ell i}|^2 |f_{\ell j}^L|^2 \nonumber\\
& &\mbox{\hspace*{-.5cm}}\Big[ (|\Delta^t(\tilde{e}_L)|^2
+|\Delta^u(\tilde{e}_L)|^2)
(E_i E_j+q^2 \cos^2\Theta)
-(|\Delta^t(\tilde{e}_L)|^2-|\Delta^u(\tilde{e}_L)|^2)
2 E_b q \cos\Theta\Big]
\nonumber\\
& &\mbox{\hspace*{-.5cm}}
-Re\{(f^{L*}_{\ell i})^2 (f^L_{\ell j})^2
     \Delta^u(\tilde{e}_L)\Delta^{t*}(\tilde{e}_L)\}
2 \eta_i \eta_j m_i m_j\Big\}.\label{eq_18}
\end{eqnarray}
$P(Z\tilde{e}_R),P(\tilde{e}_R \tilde{e}_R)$:
To obtain these quantities 
one
has to exchange in
eqs.~(\ref{eq_17}) and (\ref{eq_18})
\begin{eqnarray}
&& \phantom{P(Z\tilde{\ell}_R),P(\tilde{\ell}_R \tilde{\ell}_R):  }
\Delta^{t}(\tilde{e}_L)\to \Delta^{t}(\tilde{e}_R),\quad
\Delta^{u}(\tilde{e}_L)\to \Delta^{u}(\tilde{e}_R),\label{eq_lr1}
\nonumber\\
&& \phantom{P(Z\tilde{\ell}_R),P(\tilde{\ell}_R \tilde{\ell}_R):  }   
c_L(Z \tilde{e}_L)\to c_R(Z \tilde{e}_R),\quad 
c_L(\tilde{e}_L \tilde{e}_L)\to c_R(\tilde{e}_R \tilde{e}_R),
\label{eq_lr2}\nonumber\\
&& \phantom{P(Z\tilde{\ell}_R),P(\tilde{\ell}_R \tilde{\ell}_R):  }
O^{''L}_{ij}\to O^{''R}_{ij},\quad
 f_{\ell i}^L\to f_{\ell i}^R,\quad
f_{\ell j}^L\to f_{\ell j}^R.\nonumber\label{eq_lr3}
\end{eqnarray}
\noindent  
The propagators are defined as follows:
\begin{eqnarray}
& &\Delta^s(Z)=\frac{i}{s-m^2_Z+im_Z\Gamma_Z},
\nonumber\\
& &\Delta^{t}(\tilde{e}_{L,R})=
\frac{i}{t-m^2_{\tilde{e}_{L,R}}
+im_{\tilde{e}_{L,R}}\Gamma_{\tilde{e}_{L,R}}},\quad
\Delta^u (\tilde{e}_{L,R})=
\frac{i}{u-m^2_{\tilde{e}_{L,R}}+im_{\tilde{e}_{L,R}}
\Gamma_{\tilde{e}_{L,R}}},\label{eq_11}
\end{eqnarray}
where $m_Z, \Gamma_Z$, $m_{\tilde{e}_{L}}, \Gamma_{\tilde{e}_{L}}$,
$m_{\tilde{e}_{R}}, \Gamma_{\tilde{e}_{R}}$
denote the corresponding mass and width of the exchanged particle.

The angle $\Theta$ is the scattering angle between the 
incoming $e^{-}(p_1)$ beam and the
outgoing neutralino $\tilde{\chi}^0_j(p_4)$, the azimuth can be chosen
equal to zero. For our study of the whole process of production and
subsequent decay it is convenient to choose a coordinate frame in the 
laboratory system, where the momenta are given by: 
\begin{eqnarray}
p_1 &=& E_b(1,-\sin\Theta,0, \cos\Theta)\label{eq_19a},\\
p_2 &=& E_b(1, \sin\Theta,0,-\cos\Theta),\\
p_3 &=& (E_i,0,0,-q),\\
p_4 &=& (E_j,0,0, q),\label{eq_19}
\end{eqnarray}
with
\begin{equation}
E_i=\frac{s+m_i^2-m_j^2}{2 \sqrt{s}},\quad
E_j=\frac{s+m_j^2-m_i^2}{2 \sqrt{s}},\quad
q=\frac{\lambda^{\frac{1}{2}}(4 E_b^2,m_i^2,m_j^2)}{2 \sqrt{s}},\label{eq_20}
\end{equation}
where $m_i, m_j$ the masses of the
neutralinos and $\lambda$ the kinematical triangle function which is
given by $\lambda(x,y,z)=x^2+y^2+z^2-2xy-2xz-2yz$.
\subsection{Contributions of neutralino polarization}
Now we give the terms of eq.~(\ref{eq_14}) which depend on the 
polarization states of
the neutralinos. For the neutralino $\tilde{\chi}^0_i$$(\tilde{\chi}^0_j)$
 with momentum $p_3$$(p_4)$ we introduce 
three spacelike polarization vectors
$s^a_{\mu}(\tilde{\chi}^0_i)$($s^b_{\mu}(\tilde{\chi}^0_j)$),
(a, b=1, 2, 3), which together with $p_3^{\mu}/m$($p_4^{\mu}/m$) form an 
orthonormal set \cite{haber}.
In the laboratory system, see eqs.~(\ref{eq_19a})--(\ref{eq_19}), we choose the
following set of polarization vectors:
\begin{eqnarray}
s^1(\tilde{\chi}^0_i)&=&(0,-1,0,0),\label{eq_21}\\
s^2(\tilde{\chi}^0_i)&=&(0,0,1,0),\label{eq_22}\\
s^3(\tilde{\chi}^0_i)&=&\frac{1}{m_i}(q,0,0,-E_i),\label{eq_23}\\
& &\nonumber\\
s^1(\tilde{\chi}^0_j)&=&(0,1,0,0)\label{eq_24},\\
s^2(\tilde{\chi}^0_j)&=&(0,0,1,0)\label{eq_25},\\
s^3(\tilde{\chi}^0_j)&=&\frac{1}{m_j}(q,0,0,E_j),\label{eq_26}
\end{eqnarray}
where $s^3$ denote the longitudinal polarization, $s^1$ transverse
polarization in the scattering plane, and $s^2$ the transverse
polarization perpendicular to the scattering plane.
\subsubsection{Polarization of $\tilde{\chi}^0_i$}
We give the expression for $\Sigma^a_P(\tilde{\chi}^0_i)$ of 
eq.~(\ref{eq_14}), 
where $a=$1, 2, 3 indicates the direction of the polarization vector
$s^a(\tilde{\chi}^0_i)$, as given in eqs.~(\ref{eq_21})--(\ref{eq_23}).
It can be decomposed as:
\begin{equation}
\Sigma_P^a(\tilde{\chi}^0_i)=
 \Sigma_P^a(\tilde{\chi}^0_i,ZZ)
+\Sigma_P^a(\tilde{\chi}^0_i,Z\tilde{e}_L)
+\Sigma_P^a(\tilde{\chi}^0_i,Z\tilde{e}_R)
+\Sigma_P^a(\tilde{\chi}^0_i,\tilde{e}_L \tilde{e}_L)
+\Sigma_P^a(\tilde{\chi}^0_i,\tilde{e}_R \tilde{e}_R).\label{eq_27}
\end{equation}
\begin{enumerate}
\item The contributions of 
transverse polarization $s^1(\tilde{\chi}^0_i)$ 
{\it in the scattering plane} read:
\begin{eqnarray}
\Sigma_P^1(\tilde{\chi}^0_i,ZZ)&=&
2 \frac{g^4}{\cos^4\Theta_W}
|\Delta^s(Z)|^2 E_b^2 \sin\Theta
(c_{R}(ZZ)-c_{L}(ZZ)) \nonumber\\
& &\Big[ |O^{''L}_{ij}|^2\eta_i m_i E_j
-[(Re O^{''L}_{ij})^2 -(Im O^{''L}_{ij})^2] \eta_j m_j E_i\Big],\label{s1_zz}\\
\Sigma_P^1(\tilde{\chi}^0_i,Z\tilde{e}_L) &=& 
\frac{g^4}{\cos^2\Theta_W} c_{L}(Z \tilde{e}_L) E_b^2 \sin\Theta
\nonumber\\ 
& &
\Big[- Re\big\{\Delta^s(Z)
\big[f^{L}_{\ell i}f^{L*}_{\ell j} O^{''L*}_{ij} \Delta^{u*}(\tilde{e}_L)
+f^{L*}_{\ell i}f^{L}_{\ell j} O^{''L}_{ij} \Delta^{t*}(\tilde{e}_L)\big]
\eta_i m_i E_j\big\}\nonumber\\
& &
+Re\big\{\Delta^s(Z)
\big[f^{L}_{\ell i}f^{L*}_{\ell j} O^{''L}_{ij} 
\Delta^{u*}(\tilde{e}_L)
+f^{L*}_{\ell i}f^{L}_{\ell j} O^{''L*}_{ij} \Delta^{t*}(\tilde{e}_L)\big]
\eta_j m_j E_i\big\}\nonumber\\
& &
-Re\big\{\Delta^s(Z)
\big[f^{L}_{\ell i}f^{L*}_{\ell j} O^{''L*}_{ij} \Delta^{u*}(\tilde{e}_L)
-f^{L*}_{\ell i}f^{L}_{\ell j} O^{''L}_{ij} \Delta^{t*}(\tilde{e}_L)\big]
\eta_i m_i q \cos\Theta\big\}\Big],\nonumber\\&&\label{s1_zl}\\
\Sigma_P^1(\tilde{\chi}^0_i,\tilde{e}_L \tilde{e}_L)&=&
-\frac{g^4}{4} c_{L}(\tilde{e}_L \tilde{e}_L) E_b^2 \sin\Theta
\Big\{ |f^L_{\ell i}|^2 |f^L_{\ell j}|^2 \nonumber\\
& &\mbox{\hspace*{-.5cm}}
\Big[ (|\Delta^t (\tilde{e}_L)|^2+|\Delta^u (\tilde{e}_L)|^2)
\eta_i m_i E_j 
-(|\Delta^t (\tilde{e}_L)|^2-|\Delta^u (\tilde{e}_L)|^2)
\eta_i m_i q \cos\Theta \Big]\nonumber\\
& &\mbox{\hspace*{-.5cm}}
-2 Re\{ (f^{L*}_{\ell i})^2 (f^{L}_{\ell j})^2
\Delta^{u}(\tilde{e}_L)\Delta^{t*}(\tilde{e}_L)\}\eta_j m_j
E_i\Big\}.
\label{s1_ll}
\end{eqnarray}
$\Sigma_P^1(\tilde{\chi}^0_i,Z \tilde{e}_R),
\Sigma_P^1(\tilde{\chi}^0_i,\tilde{e}_R \tilde{e}_R)$:
To obtain these quantities one has to exchange in
eqs.(\ref{s1_zl})\\
\phantom{$\Sigma_P^1(\tilde{\chi}^0_i,Z \tilde{e}_R),
\Sigma_P^1(\tilde{\chi}^0_i,\tilde{e}_R \tilde{e}_R)$:}
and (\ref{s1_ll})
\begin{eqnarray}
& &\phantom{\Sigma_P^1(\tilde{\chi}^0_i,Z \tilde{e}_R),
\Sigma_P^1(\tilde{\chi}^0_i,\tilde{e}_R \tilde{e}_R)}
\Delta^{t}(\tilde{e}_L)\to\Delta^{t}(\tilde{e}_R),\quad
\Delta^{u}(\tilde{e}_L)\to\Delta^{u}(\tilde{e}_R),
\label{lr_4}\nonumber\\
& &\phantom{\Sigma_P^1(\tilde{\chi}^0_i,Z \tilde{e}_R),
\Sigma_P^1(\tilde{\chi}^0_i,\tilde{e}_R \tilde{e}_R)}
c_L(Z \tilde{e}_L)\to c_R(Z \tilde{e}_R),\quad 
c_L(\tilde{e}_L \tilde{e}_L)\to c_R(\tilde{e}_R \tilde{e}_R),
\label{le_5}\nonumber\\
& &\phantom{\Sigma_P^1(\tilde{\chi}^0_i,Z \tilde{e}_R),
 \Sigma_P^1(\tilde{\chi}^0_i,\tilde{e}_R \tilde{e}_R)}
O^{''L}_{ij}\to O^{''R}_{ij},\quad
 f_{\ell i}^L\to f_{\ell i}^R,\quad
 f_{\ell j}^L\to f_{\ell j}^R,\nonumber\label{lr_6}
\end{eqnarray}
\phantom{$\Sigma_P^1(\tilde{\chi}^0_i,Z \tilde{e}_R),
\Sigma_P^1(\tilde{\chi}^0_i,\tilde{e}_R \tilde{e}_R)$:} and to change
 the overall sign of the right hand side of\\ 
\phantom{$\Sigma_P^1(\tilde{\chi}^0_i,Z \tilde{e}_R),
\Sigma_P^1(\tilde{\chi}^0_i,\tilde{e}_R \tilde{e}_R)$:}
eqs.~(\ref{s1_zl}), (\ref{s1_ll}). 
\item The contributions of longitudinal polarization  
$s^3(\tilde{\chi}^0_i)$ read:
\begin{eqnarray}
\Sigma_P^3(\tilde{\chi}^0_i,ZZ)&=&
\eta_i  \frac{2 g^4}{\cos^4\Theta_W}
|\Delta^s(Z)|^2  (c_{L}(ZZ)-c_{R}(ZZ)) E_b^2 \cos\Theta
\nonumber\\
& &
\Big[ |O^{''L}_{ij}|^2 (E_i E_j+q^2)
-[(Re O^{''L}_{ij})^2-(Im O^{''L*}_{ij})^2]\eta_i\eta_j m_i  m_j
\Big],\label{s3_zz}\\
\Sigma_P^3(\tilde{\chi}^0_i,Z\tilde{e}_L) &=& \eta_i
\frac{g^4}{\cos^2\Theta_W} c_{L}(Z \tilde{e}_L) 
E_b^2 
\nonumber\\ & &\mbox{\hspace*{-1cm}}
\Big[ Re\Big\{ \Delta^s(Z) 
[f^{L}_{\ell i}f^{L*}_{\ell j} O^{''L*}_{ij} \Delta^{u*}(\tilde{e}_L)
 -f^{L*}_{\ell i}f^{L}_{\ell j} O^{''L}_{ij} 
\Delta^{t*}(\tilde{e}_L)]
 E_j q\Big\}\nonumber\\
& &\mbox{\hspace*{-1.2cm}}
+Re\Big\{ \Delta^s(Z) 
[f^{L}_{\ell i}f^{L*}_{\ell j} O^{''L*}_{ij} \Delta^{u*}(\tilde{e}_L)
 +f^{L*}_{\ell i}f^{L}_{\ell j} O^{''L}_{ij} \Delta^{t*}(\tilde{e}_L)]
 (E_i E_j+ q^2)\cos\Theta\Big\}\nonumber\\
& &\mbox{\hspace*{-1.2cm}}
+Re\Big\{ \Delta^s(Z) 
[f^{L}_{\ell i}f^{L*}_{\ell j} O^{''L*}_{ij} \Delta^{u*}(\tilde{e}_L)
 -f^{L*}_{\ell i}f^{L}_{\ell j} O^{''L}_{ij} \Delta^{t*}(\tilde{e}_L)]
 E_i q \cos^2\Theta\Big\}\nonumber\\
& &\mbox{\hspace*{-1.2cm}}
-Re\Big\{ \Delta^s(Z) 
[f^{L}_{\ell i}f^{L*}_{\ell j} O^{''L}_{ij} \Delta^{u*}(\tilde{e}_L)
 +f^{L*}_{\ell i}f^{L}_{\ell j} O^{''L*}_{ij} \Delta^{t*}(\tilde{e}_L)]
\eta_i \eta_j m_i m_j \cos\Theta\Big\}\Big],\nonumber\\&&\label{s3_zl}\\
\Sigma_P^3(\tilde{\chi}^0_i,\tilde{e}_L \tilde{e}_L)&=&
\eta_i \frac{g^4}{4} c_{L}(\tilde{e}_L \tilde{e}_L) 
E_b^2
\Big[ |f^L_{\ell i}|^2 |f^L_{\ell j}|^2 \nonumber\\&&
\Big\{ [|\Delta^u (\tilde{e}_L)|^2-|\Delta^t (\tilde{e}_L)|^2]
E_j q+[|\Delta^u (\tilde{e}_L)|^2
-|\Delta^t (\tilde{e}_L)|^2]
q E_i  \cos^2 \Theta
\nonumber\\
& &
+[|\Delta^t (\tilde{e}_L)|^2+|\Delta^u (\tilde{e}_L)|^2]
(E_i E_j +q^2) \cos\Theta
\Big\}\nonumber\\
& &
-2 Re\{(f^{L*}_{\ell i})^2 (f^{L}_{\ell j})^2
\Delta^u (\tilde{e}_L)\Delta^{t*} (\tilde{e}_L)\}
\eta_i \eta_j m_i m_j \cos\Theta\Big].\label{s3_ll}
\end{eqnarray}
$\Sigma_P^3(\tilde{\chi}^0_i,Z \tilde{e}_R),
\Sigma_P^3(\tilde{\chi}^0_i,\tilde{e}_R \tilde{e}_R)$:
To obtain these quantities
one has
to exchange in eqs.~(\ref{s3_zl})\\
\phantom{$\Sigma_P^3(\tilde{\chi}^0_i,Z \tilde{e}_R),
\Sigma_P^3(\tilde{\chi}^0_i,\tilde{e}_R \tilde{e}_R)$:}
and (\ref{s3_ll}) 
\begin{eqnarray}
&&\phantom{\Sigma_P^3(\tilde{\chi}^0_i,Z \tilde{e}_R),
\Sigma_P^3(\tilde{\chi}^0_i,\tilde{e}_R \tilde{e}_R)}
\Delta^{t}(\tilde{e}_L)\to\Delta^{t}(\tilde{e}_R),\quad
\Delta^{u}(\tilde{e}_L)\to\Delta^{u}(\tilde{e}_R),
\label{lr_7}\nonumber\\
&&\phantom{\Sigma_P^3(\tilde{\chi}^0_i,Z \tilde{e}_R),
\Sigma_P^3(\tilde{\chi}^0_i,\tilde{e}_R \tilde{e}_R)}
c_L(Z \tilde{e}_L)\to c_R(Z \tilde{e}_R),\quad 
c_L(\tilde{e}_L \tilde{e}_L)\to c_R(\tilde{e}_R \tilde{e}_R),
\label{lr_8}\nonumber\\
&&\phantom{\Sigma_P^3(\tilde{\chi}^0_i,Z \tilde{e}_R),
\Sigma_P^3(\tilde{\chi}^0_i,\tilde{e}_R \tilde{e}_R)}
O^{''L}_{ij}\to O^{''R}_{ij},\quad
 f_{\ell i}^L\to f_{\ell i}^R,\quad
 f_{\ell j}^L\to f_{\ell j}^R. \nonumber\label{lr_9}
\end{eqnarray}
\phantom{$\Sigma_P^3(\tilde{\chi}^0_i,Z \tilde{e}_R),
\Sigma_P^3(\tilde{\chi}^0_i,\tilde{e}_R \tilde{e}_R)$}
and to change the overall sign of 
eqs.~(\ref{s3_zl}), (\ref{s3_ll}).
\item The contributions of the polarization
$s^2(\tilde{\chi}^0_i)$ perpendicular {\it to the scattering plane} are:
\begin{eqnarray}
\Sigma_P^2(\tilde{\chi}^0_i,ZZ)&=&
-4 (\frac{g^2}{\cos^2\Theta_W})^2
|\Delta^s(Z)|^2  (c_{R}(ZZ)-c_{L}(ZZ)) \nonumber\\
& & m_j q E_b^2 
\sin\Theta Re(O^{''L}_{ij}) Im(O^{''L}_{ij}),\label{s2_zz}\\
\Sigma_P^2(\tilde{\chi}^0_i,Z \tilde{e}_L) &=& 
\frac{g^4}{\cos^2\Theta_W} c_{L}(Z \tilde{e}_L) \eta_j m_j
E_b^2 q \sin\Theta \nonumber\\
& & Im\Big\{\Delta^s(Z)
 \big[f^{L}_{\ell i}f^{L*}_{\ell j} O^{''L}_{ij} 
\Delta^{u*}(\tilde{e}_L) -f^{L*}_{\ell i}f^{L}_{\ell j} O^{''L*}_{ij} \Delta^{t*}(\tilde{e}_L) \big]\Big\},\nonumber\\ & &\label{s2_zl}\\
\Sigma_P^2(\tilde{\chi}^0_i,\tilde{e}_L \tilde{e}_L)&=& 
-\frac{g^4}{2} c_{L}(\tilde{e}_L \tilde{e}_L) 
\eta_j m_j E_b^2 q \sin\Theta 
Im\Big\{(f^{L*}_{\ell i})^2 (f^{L}_{\ell j})^2
\Delta^u (\tilde{e}_L)\Delta^{t*} (\tilde{e}_L)\Big\}.\label{s2_ll}
\end{eqnarray}
$\Sigma_P^2(\tilde{\chi}^0_i,Z \tilde{e}_L),
\Sigma_P^2(\tilde{\chi}^0_i,\tilde{e}_L \tilde{e}_L):$
To obtain these quantities
one has to exchange in\\
\phantom{$\Sigma_P^2(\tilde{\chi}^0_i,Z \tilde{e}_L),
\Sigma_P^2(\tilde{\chi}^0_i,\tilde{e}_L \tilde{e}_L):$}
eq.~(\ref{s2_zl}) and eq.~(\ref{s2_ll})
\begin{eqnarray}
&&\phantom{\Sigma_P^2(\tilde{\chi}^0_i,Z \tilde{\ell}_L),
\Sigma_P^2(\tilde{\chi}^0_i,\tilde{\ell}_L \tilde{\ell}_L):}
\Delta^{t}(\tilde{e}_L)\to\Delta^{t}(\tilde{e}_R),\quad
\Delta^{u}(\tilde{e}_L)\to\Delta^{u}(\tilde{e}_R),
\label{lr_10}\nonumber\\
&&\phantom{\Sigma_P^2(\tilde{\chi}^0_i,Z \tilde{\ell}_L),
\Sigma_P^2(\tilde{\chi}^0_i,\tilde{\ell}_L \tilde{\ell}_L):}
c_L(Z \tilde{e}_L)\to c_R(Z \tilde{e}_R),\quad 
c_L(\tilde{e}_L \tilde{e}_L)\to c_R(\tilde{e}_R \tilde{e}_R),
\label{lr_11}\nonumber\\
&&\phantom{\Sigma_P^2(\tilde{\chi}^0_i,Z \tilde{\ell}_L),
\Sigma_P^2(\tilde{\chi}^0_i,\tilde{\ell}_L \tilde{\ell}_L):}
O^{''L}_{ij}\to O^{''R}_{ij},\quad
 f_{\ell i}^L\to f_{\ell i}^R,\quad
 f_{\ell j}^L\to f_{\ell j}^R.\nonumber\label{lr_12}
\end{eqnarray}
\end{enumerate}
Contrary to the case of $s^1(\tilde{\chi}^0_i)$ and 
$s^3(\tilde{\chi}^0_i)$ the
sign of the contributions $\Sigma^2_P(\tilde{\chi}^0_i)$ does not
change when going from $\tilde{e}_L$ exchange to  
$\tilde{e}_R$ exchange.
\subsubsection{Polarization of $\tilde{\chi}^0_j$}
We give the quantities $\Sigma^b_P(\tilde{\chi}^0_j)$ of eq.~(\ref{eq_14}) 
which contain only the polarization
vector $s^b(\tilde{\chi}^{0}_j)$ with $b=$1, 2, 3, 
eqs.~(\ref{eq_24})--(\ref{eq_26}):
\begin{equation}
\Sigma_P^b(\tilde{\chi}^0_j)=
 \Sigma_P^b(\tilde{\chi}^0_j,ZZ)
+\Sigma_P^b(\tilde{\chi}^0_j,Z\tilde{e}_L)
+\Sigma_P^b(\tilde{\chi}^0_j,Z\tilde{e}_R)
+\Sigma_P^b(\tilde{\chi}^0_j,\tilde{e}_L \tilde{e}_L)
+\Sigma_P^b(\tilde{\chi}^0_j,\tilde{e}_R
\tilde{e}_R).\label{eq_28}
\end{equation}
\begin{enumerate}
\item
{\bf $\Sigma_P^{1}(\tilde{\chi}^0_j)$,
  $\Sigma_P^{3}(\tilde{\chi}^0_j)$}: The expressions for 
$\Sigma_P^{1}(\tilde{\chi}^0_j)$, $\Sigma_P^{3}(\tilde{\chi}^0_j)$
 are obtained
from those of
$\Sigma_P^{1}(\tilde{\chi}^0_i)$,
$\Sigma_P^{3}(\tilde{\chi}^0_i)$, 
eqs.~(\ref{s1_zz})--(\ref{s3_ll}), by exchanging 
\begin{equation}
m_i\to m_j,\quad \eta_i\to\eta_j,\quad E_i\to E_j,\label{eq_j1}
\end{equation} 
and by changing the overall sign of
these expressions, for example,
\begin{eqnarray}
\Sigma_P^1(\tilde{\chi}^0_j,ZZ)&=&
-2 \frac{g^4}{\cos^4\Theta_W}
|\Delta(Z)|^2 E_b^2 \sin\Theta
(c_{R}(ZZ)-c_{L}(ZZ)) \nonumber\\
& &\Big[ |O^{''L}_{ij}|^2\eta_j m_j E_i
-[(Re O^{''L}_{ij})^2 -(Im O^{''L}_{ij})^2] \eta_i m_i E_j\Big].
\label{sj1_zz}
\end{eqnarray}
\item
{\bf $\Sigma^2_P(\tilde{\chi}^0_j)$}: The expressions for 
$\Sigma^2_P(\tilde{\chi}^0_j)$ are obtained from
those for $\Sigma^2_P(\tilde{\chi}^0_i)$, eqs.~(\ref{s2_zz})--(\ref{s2_ll}) 
by exchanging 
\begin{equation}
\eta_i\to\eta_j,\quad m_i\to m_j, \quad E_i\to E_j\label{eq_j2}
\quad\mbox{(without changing the overall sign)}.
\end{equation}
\end{enumerate}

\noindent Note that
\begin{itemize}
\item the transverse polarizations 
$\Sigma^1_P(\tilde{\chi}^0_i)$, $\Sigma^2_P(\tilde{\chi}^0_i)$,
$\Sigma^1_P(\tilde{\chi}^0_j)$, $\Sigma^2_P(\tilde{\chi}^0_j)$
of the neutralinos vanish in forward and backward direction;
\item at threshold the tranverse polarizations 
$\Sigma^2_P(\tilde{\chi}^0_i)$ and
$\Sigma^2_P(\tilde{\chi}^0_j)$ perpendicular to the production plane 
vanish proportional to the momentum of the neutralinos.
\end{itemize}
\subsubsection{Spin-spin correlations}
We give the expressions for $\Sigma^{ab}_P(\tilde{\chi}^0_i
\tilde{\chi}^0_j)$ 
of eq.~(\ref{eq_14}), 
where $a$, $b=$1, 2, 3 indicate the directions of the polarization vectors
$s^a(\tilde{\chi}^0_i)$ and $s^b(\tilde{\chi}^0_j)$ as given in 
eqs.~(\ref{eq_21})--(\ref{eq_26}). 
They can be decomposed as:
\begin{eqnarray}
\Sigma_P^{ab}(\tilde{\chi}^0_i\tilde{\chi}^0_j)&=&
 \Sigma_P^{ab}(\tilde{\chi}^0_i\tilde{\chi}^0_j,ZZ)
+\Sigma_P^{ab}(\tilde{\chi}^0_i\tilde{\chi}^0_j,Z\tilde{e}_L)
+\Sigma_P^{ab}(\tilde{\chi}^0_i\tilde{\chi}^0_j,Z\tilde{e}_R)
+\Sigma_P^{ab}(\tilde{\chi}^0_i\tilde{\chi}^0_j,\tilde{e}_L \tilde{e}_L)
\nonumber\\& &
+\Sigma_P^{ab}(\tilde{\chi}^0_i\tilde{\chi}^0_j,\tilde{e}_R 
\tilde{e}_R), \quad\mbox{with }a\mbox{, }b=1\mbox{, }2\mbox{, }3. 
\label{eq_29}
\end{eqnarray}
\begin{enumerate}
\item {\sf The contributions of $s^1(\tilde{\chi}^0_i)$ and 
$s^1(\tilde{\chi}^0_j)$ are:}
\begin{eqnarray}
\Sigma_P^{11}(\tilde{\chi}^0_i\tilde{\chi}^0_j,ZZ)&=&
2 \frac{g^4}{\cos^4\Theta_W}
|\Delta^s(Z)|^2 (c_{R}(ZZ)+c_{L}(ZZ))
E_b^2 \sin^2\Theta
\nonumber\\&&
\big\{[(Re O^{''L}_{ij})^2- (Im O^{''L}_{ij})^2]
E_i E_j -2 |O^{''L}_{ij}|^2 \eta_i \eta_j m_i m_j\big\},\label{ss11_zz}\\
\Sigma_P^{11}(\tilde{\chi}^0_i\tilde{\chi}^0_j,
Z \tilde{e}_L)&=& \frac{g^4}{\cos^2\Theta_W}
c_{L}(Z \tilde{e}_L) E_b^2 \sin^2\Theta
\nonumber\\& &\mbox{\hspace*{-.8cm}}
Re\Big\{
[f^{L}_{\ell i}f^{L*}_{\ell j}\Delta^s(Z)\Delta^{u*}(\tilde{e}_L)
O^{''L}_{ij}
+f^{L*}_{\ell i}f^{L}_{\ell j}\Delta^s(Z)\Delta^{t*}(\tilde{e}_L)
O^{''L*}_{ij}]E_i E_j\nonumber\\ 
& &\mbox{\hspace*{-.8cm}}
-[f^{L}_{\ell i}f^{L*}_{\ell j}\Delta^s(Z)\Delta^{u*}(\tilde{e}_L)
O^{''L*}_{ij}
+f^{L*}_{\ell i}f^{L}_{\ell j}\Delta^s(Z)\Delta^{t*}(\tilde{e}_L)
O^{''L}_{ij}]\eta_i \eta_j m_i m_j\Big\},\nonumber\\&& \label{ss11_zl}\\
\Sigma_P^{11}(\tilde{\chi}^0_i\tilde{\chi}^0_j,
\tilde{e}_L \tilde{e}_L)&=&
-\frac{g^4}{4} c_{L}(\tilde{e}_L\tilde{e}_L)E_b^2 \sin^2\Theta
\nonumber\\
&&\Big[|f^L_{\ell i}|^2 |f^L_{\ell j}|^2 
(|\Delta^t(\tilde{e}_L)|^2
+|\Delta^u(\tilde{e}_L)|^2)\eta_i \eta_j m_i m_j\nonumber\\
& &
+2 Re\big\{ (f^{L*}_{\ell i})^2 (f^{L}_{\ell j})^2
\Delta^u(\tilde{e}_L)\Delta^{t*}(\tilde{e}_L)\big\}
E_i E_j\Big].\label{ss11_ll}
\end{eqnarray}
$\Sigma_P^{11}(\tilde{\chi}^0_i\tilde{\chi}^0_j,
Z \tilde{e}_R),
\Sigma_P^{11}(\tilde{\chi}^0_i\tilde{\chi}^0_j,
\tilde{e}_R \tilde{e}_R)$:
To obtain these quantities
one has to exchange\\  
\phantom{$\Sigma_P^{11}(\tilde{\chi}^0_i\tilde{\chi}^0_j,
Z \tilde{\ell}_R),
\Sigma_P^{11}(\tilde{\chi}^0_i\tilde{\chi}^0_j,
\tilde{\ell}_R \tilde{\ell}_R)$: } 
in eqs.~(\ref{ss11_zl}) and (\ref{ss11_ll})
\begin{eqnarray}
&&\phantom{\Sigma_P^{11}(\tilde{\chi}^0_i\tilde{\chi}^0_j,
Z \tilde{\ell}_R),
\Sigma_P^{11}(\tilde{\chi}^0_i\tilde{\chi}^0_j,
\tilde{\ell}_R \tilde{\ell}_R)}
\Delta^{t}(\tilde{e}_L)\to\Delta^{t}(\tilde{e}_R),\quad
\Delta^{u}(\tilde{e}_L)\to\Delta^{u}(\tilde{e}_R),
\label{lr_13}\nonumber\\
& &\phantom{\Sigma_P^{11}(\tilde{\chi}^0_i\tilde{\chi}^0_j,
Z \tilde{\ell}_R),\Sigma_P^{11}(\tilde{\chi}^0_i\tilde{\chi}^0_j,
\tilde{\ell}_R\tilde{\ell}_R)}
c_L(Z \tilde{e}_L)\to c_R(Z \tilde{e}_R),\quad 
c_L(\tilde{e}_L \tilde{e}_L)\to c_R(\tilde{e}_R \tilde{e}_R),
\label{lr_14}\nonumber\\
& &\phantom{\Sigma_P^{11}(\tilde{\chi}^0_i\tilde{\chi}^0_j,
Z \tilde{\ell}_R),\Sigma_P^{11}(\tilde{\chi}^0_i\tilde{\chi}^0_j,
\tilde{\ell}_R\tilde{\ell}_R)}
O^{''L}_{ij}\to O^{''R}_{ij},\quad
 f_{\ell i}^L\to f_{\ell i}^R,\quad
 f_{\ell j}^L\to f_{\ell j}^R.\nonumber\label{lr_15}
\end{eqnarray}
\item {\sf The contributions of $s^2(\tilde{\chi}^0_i)$ and 
$s^2(\tilde{\chi}^0_j)$ are:}
\begin{eqnarray}
\Sigma_P^{22}(\tilde{\chi}^0_i\tilde{\chi}^0_j,ZZ)&=&
2 \frac{g^4}{\cos^4\Theta_W}
|\Delta^s(Z)|^2 (c_{R}(ZZ)+c_{L}(ZZ)) E_b^2 q^2 \sin^2\Theta 
\nonumber\\&&
\{(Re O^{''L}_{ij})^2-(Im O^{''L}_{ij})^2\}, \label{ss22_zz} \\
\Sigma_P^{22}(\tilde{\chi}^0_i\tilde{\chi}^0_j,Z \tilde{e}_L)&=&
\frac{g^4}{\cos^2\Theta_W}c_{L}(Z \tilde{e}_L)
E_b^2 q^2 \sin^2\Theta\nonumber\\&&
Re\Big\{\Delta^s(Z)[
f^{L}_{\ell i}f^{L*}_{\ell j}\Delta^{u*}(\tilde{e}_L) O^{''L}_{ij}
+f^{L*}_{\ell i}f^{L}_{\ell j}\Delta^{t*}(\tilde{e}_L) 
O^{''L*}_{ij}]\Big\},\label{ss22_zl}\\
\Sigma_P^{22}(\tilde{\chi}^0_i\tilde{\chi}^0_j,
\tilde{e}_L \tilde{e}_L)&=&
-\frac{g^4}{2} c_{L}(\tilde{e}_L\tilde{e}_L)
E_b^2 q^2 \sin^2\Theta
Re\Big\{(f^{L*}_{\ell i})^2 (f^{L}_{\ell j})^2 \Delta^u(\tilde{e}_L)
\Delta^{t*}(\tilde{e}_L)\Big\}\label{ss22_ll}.
\end{eqnarray}
$\Sigma_P^{22}(\tilde{\chi}^0_i\tilde{\chi}^0_j,Z \tilde{e}_R),
\Sigma_P^{22}(\tilde{\chi}^0_i\tilde{\chi}^0_j,
\tilde{e}_R \tilde{e}_R):$
To obtain these quantities
one has to exchange\\
\phantom{$\Sigma_P^{22}(\tilde{\chi}^0_i\tilde{\chi}^0_j,Z \tilde{e}_R),
\Sigma_P^{22}(\tilde{\chi}^0_i\tilde{\chi}^0_j,
\tilde{e}_R \tilde{e}_R):$ }
in eqs.~(\ref{ss22_zl})
and (\ref{ss22_ll})
\begin{eqnarray}
&&\phantom{\Sigma_P^{22}(\tilde{\chi}^0_i\tilde{\chi}^0_j,Z \tilde{e}_R),
\Sigma_P^{22}(\tilde{\chi}^0_i\tilde{\chi}^0_j,
\tilde{e}_R \tilde{e}_R)}
\Delta^{t}(\tilde{e}_L)\to\Delta^{t}(\tilde{e}_R),\quad
\Delta^{u}(\tilde{e}_L)\to\Delta^{u}(\tilde{e}_R),
\label{lr_16}\nonumber\\
&&\phantom{\Sigma_P^{22}(\tilde{\chi}^0_i\tilde{\chi}^0_j,Z \tilde{e}_R),
\Sigma_P^{22}(\tilde{\chi}^0_i\tilde{\chi}^0_j,
\tilde{e}_R \tilde{e}_R)}
c_L(Z \tilde{e}_L)\to c_R(Z \tilde{e}_R),\quad 
c_L(\tilde{e}_L \tilde{e}_L)\to c_R(\tilde{e}_R \tilde{e}_R),
\label{lr_17}\nonumber\\
&&\phantom{\Sigma_P^{22}(\tilde{\chi}^0_i\tilde{\chi}^0_j,Z \tilde{e}_R),
\Sigma_P^{22}(\tilde{\chi}^0_i\tilde{\chi}^0_j,
\tilde{e}_R \tilde{e}_R)}
O^{''L}_{ij}\to O^{''R}_{ij},\quad
 f_{\ell i}^L\to f_{\ell i}^R,\quad
 f_{\ell j}^L\to f_{\ell j}^R.\label{lr_18}\nonumber
\end{eqnarray}
\item {\sf The contributions of $s^3(\tilde{\chi}^0_i)$ and 
$s^3(\tilde{\chi}^0_j)$ are:}
\begin{eqnarray}
\Sigma_P^{33}(\tilde{\chi}^0_i\tilde{\chi}^0_j,ZZ)&=&
\eta_i\eta_j  \frac{2 g^4}{\cos^4\Theta_W}
|\Delta^s(Z)|^2 (c_{R}(ZZ)+c_{L}(ZZ)) E_b^2
\nonumber\\&&
\Big[ ((Re O^{''L}_{ij})^2-(Im O^{''L}_{ij})^2) 
\eta_i\eta_j m_i m_j \cos^2\Theta
\nonumber\\&&
-|O^{''L}_{ij}|^2
  [q^2+E_i E_j\cos^2\Theta]\Big],\label{ss33_zz}\\
\Sigma_P^{33}(\tilde{\chi}^0_i\tilde{\chi}^0_j,Z \tilde{e}_L)
&=&
\eta_i \eta_j \frac{g^4}{\cos^2\Theta_W}
c_{L}(Z \tilde{e}_L) E_b^2
\nonumber\\
& &\mbox{\hspace*{-1.2cm}} 
\Big[
Re\big\{\Delta^s(Z)
[f^{L}_{\ell i}f^{L*}_{\ell j}\Delta^{u*}(\tilde{e}_L)
O^{''L}_{ij}
+f^{L*}_{\ell i}f^{L}_{\ell j}\Delta^{t*}(\tilde{e}_L)
O^{''L*}_{ij}]\big\}\eta_i \eta_j m_i m_j \cos^2\Theta\nonumber\\ 
& &\mbox{\hspace*{-1.8cm}}
-Re\big\{\Delta^s(Z)
[f^{L}_{\ell i}f^{L*}_{\ell j}\Delta^{u*}(\tilde{e}_L)
O^{''L*}_{ij}
+f^{L*}_{\ell i}f^{L}_{\ell j}\Delta^{t*}(\tilde{e}_L)
O^{''L}_{ij}]\big\}[q^2+E_i E_j \cos^2\Theta]\nonumber\\
& &\mbox{\hspace*{-1.8cm}}
-Re\big\{\Delta^s(Z)
[f^{L}_{\ell i}f^{L*}_{\ell j}\Delta^{u*}(\tilde{e}_L)
O^{''L*}_{ij}
-f^{L*}_{\ell i}f^{L}_{\ell j}\Delta^{t*}(\tilde{e}_L)
O^{''L}_{ij}]\big\} 2 E_b q \cos\Theta \Big],\label{ss33_zl}\\
\Sigma_P^{33}(\tilde{\chi}^0_i\tilde{\chi}^0_j,\tilde{e}_L
\tilde{e}_L)&=& \eta_i\eta_j
\frac{g^4}{4} c_{L}(\tilde{e}_L\tilde{e}_L)E_b^2 
\Big[|f^L_{\ell i}|^2 |f^L_{\ell j}|^2 
\nonumber\\ 
& &\Big(
-(|\Delta^t(\tilde{e}_L)|^2+|\Delta^u(\tilde{e}_L)|^2)
[q^2+E_i E_j \cos^2\Theta]\nonumber\\
& &\phantom{\Big()}
+(|\Delta^t(\tilde{e}_L)|^2-|\Delta^u(\tilde{e}_L)|^2)
2 E_b q \cos\Theta\Big)\nonumber\\
& &
-2 Re\big\{ (f^{L*}_{\ell i})^2 (f^{L}_{\ell j})^2
\Delta^u(\tilde{e}_L)\Delta^{t*}(\tilde{e}_L)\big\}
\eta_i \eta_j m_i m_j \cos^2\Theta\Big]\label{ss33_ll}.
\end{eqnarray}
$\Sigma_P^{33}(\tilde{\chi}^0_i\tilde{\chi}^0_j,Z
\tilde{e}_R),
\Sigma_P^{33}(\tilde{\chi}^0_i\tilde{\chi}^0_j,\tilde{e}_R
\tilde{e}_R):$
To obtain these quantities
one has to exchange\\
\phantom{$\Sigma_P^{33}(\tilde{\chi}^0_i\tilde{\chi}^0_j,Z
\tilde{e}_R),
\Sigma_P^{33}(\tilde{\chi}^0_i\tilde{\chi}^0_j,\tilde{e}_R
\tilde{e}_R):$ } 
in eqs.~(\ref{ss33_zl}) and (\ref{ss33_ll})
\begin{eqnarray}
&&\phantom{\Sigma_P^{33}(\tilde{\chi}^0_i\tilde{\chi}^0_j,Z
\tilde{e}_R),
\Sigma_P^{33}(\tilde{\chi}^0_i\tilde{\chi}^0_j,\tilde{e}_R
\tilde{e}_R)}
\Delta^{t}(\tilde{e}_L)\to\Delta^{t}(\tilde{e}_R),\quad
\Delta^{u}(\tilde{e}_L)\to\Delta^{u}(\tilde{e}_R),
\label{lr_19}\nonumber\\
&&\phantom{\Sigma_P^{33}(\tilde{\chi}^0_i\tilde{\chi}^0_j,Z
\tilde{e}_R),
\Sigma_P^{33}(\tilde{\chi}^0_i\tilde{\chi}^0_j,\tilde{e}_R
\tilde{e}_R)}
c_L(Z \tilde{e}_L)\to c_R(Z \tilde{e}_R),\quad 
c_L(\tilde{e}_L \tilde{e}_L)\to c_R(\tilde{e}_R \tilde{e}_R),
\label{lr_20}\nonumber\\
&&\phantom{\Sigma_P^{33}(\tilde{\chi}^0_i\tilde{\chi}^0_j,Z
\tilde{e}_R),
\Sigma_P^{33}(\tilde{\chi}^0_i\tilde{\chi}^0_j,\tilde{e}_R
\tilde{e}_R)}
O^{''L}_{ij}\to O^{''R}_{ij},\quad
 f_{\ell i}^L\to f_{\ell i}^R,\quad
 f_{\ell j}^L\to f_{\ell j}^R.\label{lr_21}\nonumber
\end{eqnarray}
\item {\sf The contributions of $s^1(\tilde{\chi}^0_i)$ and 
$s^3(\tilde{\chi}^0_j)$ are}:
\begin{eqnarray}
\Sigma_P^{13}(\tilde{\chi}^0_i\tilde{\chi}^0_j,ZZ)&=&
\eta_j \frac{2 g^4}{\cos^4\Theta_W}
|\Delta^s(Z)|^2 (c_{R}(ZZ)+c_{L}(ZZ)) E_b^2 \sin\Theta\cos\Theta
\nonumber\\ & &
\Big[ -((Re O^{''L}_{ij})^2-(Im O^{''L}_{ij})^2) E_i \eta_j m_j
+ |O^{''L}_{ij}|^2 \eta_i m_i 
E_j\Big],\label{ss13_zz}\\
\Sigma_P^{13}(\tilde{\chi}^0_i\tilde{\chi}^0_j,Z \tilde{e}_L)&=&
\eta_j \frac{g^4}{\cos^2\Theta_W}c_{L}(Z \tilde{e}_L) 
E_b^2 \sin\Theta 
\nonumber\\& & \mbox{\hspace*{-1.2cm}}
\Big[
-Re\big\{\Delta^s(Z)[
f^{L}_{\ell i}f^{L*}_{\ell j}\Delta^{u*}(\tilde{e}_L) O^{''L}_{ij}
+f^{L*}_{\ell i}f^{L}_{\ell j}\Delta^{t*}(\tilde{e}_L) 
O^{''L*}_{ij}]
\big\}E_i \eta_j m_j \cos\Theta\nonumber\\
& &\mbox{\hspace*{-1cm}}
+Re\big\{\Delta^s(Z)[
f^{L}_{\ell i}f^{L*}_{\ell j}\Delta^{u*}(\tilde{e}_L) O^{''L*}_{ij}
+f^{L*}_{\ell i}f^{L}_{\ell j}\Delta^{t*}(\tilde{e}_L) 
O^{''L}_{ij}]\big\}\eta_i m_i  E_j \cos\Theta\nonumber\\
& &\mbox{\hspace*{-1cm}}
+Re\big\{\Delta^s(Z)[
f^{L}_{\ell i}f^{L*}_{\ell j}\Delta^{u*}(\tilde{e}_L) O^{''L*}_{ij}
-f^{L*}_{\ell i}f^{L}_{\ell j}\Delta^{t*}(\tilde{e}_L) 
O^{''L}_{ij}]\big\}\eta_i m_i q \Big],\label{ss13_zl}\\
\Sigma_P^{13}(\tilde{\chi}^0_i\tilde{\chi}^0_j,\tilde{e}_L
\tilde{e}_L)&=&
\eta_j \frac{g^4}{4} c_{L}(\tilde{e}_L\tilde{e}_L) 
E_b^2  \sin\Theta \nonumber\\ 
& &
\Big[|f^{L}_{\ell i}|^2 |f^{L}_{\ell j}|^2
\big\{[|\Delta^{u}(\tilde{e}_L)|^2-|\Delta^{t}(\tilde{e}_L)|^2]
\eta_i m_i q
\nonumber\\& &
+[|\Delta^{t}(\tilde{e}_L)|^2+|\Delta^{u}(\tilde{e}_L)|^2]
\eta_i m_i E_j \cos\Theta\big\}\nonumber\\ 
& &
+2 Re\big\{(f^{L*}_{\ell i})^2 (f^{L}_{\ell j})^2 \Delta^u(\tilde{e}_L)
\Delta^{t*}(\tilde{e}_L)\big\}E_i \eta_j
m_j\cos\Theta\Big]\label{ss13_ll}.
\end{eqnarray}
$\Sigma_P^{13}(\tilde{\chi}^0_i\tilde{\chi}^0_j,Z \tilde{e}_R),
\Sigma_P^{13}(\tilde{\chi}^0_i\tilde{\chi}^0_j,\tilde{e}_R\tilde{e}_R):$
To obtain these quantities
one has to exchange\\
\phantom{$\Sigma_P^{13}(\tilde{\chi}^0_i\tilde{\chi}^0_j,Z \tilde{\ell}_R),
\Sigma_P^{13}(\tilde{\chi}^0_i\tilde{\chi}^0_j,
\tilde{\ell}_R\tilde{\ell}_R):$ } 
in eqs.~(\ref{ss13_zl}) and (\ref{ss13_ll})
\begin{eqnarray}
&&\phantom{\Sigma_P^{13}(\tilde{\chi}^0_i\tilde{\chi}^0_j,Z \tilde{\ell}_R),
\Sigma_P^{13}(\tilde{\chi}^0_i\tilde{\chi}^0_j,
\tilde{\ell}_R\tilde{\ell}_R)}
\Delta^{t}(\tilde{e}_L)\to\Delta^{t}(\tilde{e}_R),\quad
\Delta^{u}(\tilde{e}_L)\to\Delta^{u}(\tilde{e}_R)
\label{lr_22}\nonumber\\
&&\phantom{\Sigma_P^{13}(\tilde{\chi}^0_i\tilde{\chi}^0_j,Z \tilde{\ell}_R),
\Sigma_P^{13}(\tilde{\chi}^0_i\tilde{\chi}^0_j,
\tilde{\ell}_R\tilde{\ell}_R)}
c_L(Z \tilde{e}_L)\to c_R(Z \tilde{e}_R),\quad 
c_L(\tilde{e}_L \tilde{e}_L)\to c_R(\tilde{e}_R \tilde{e}_R),
\label{lr_23}\nonumber\\
&&\phantom{\Sigma_P^{13}(\tilde{\chi}^0_i\tilde{\chi}^0_j,Z \tilde{\ell}_R),
\Sigma_P^{13}(\tilde{\chi}^0_i\tilde{\chi}^0_j,
\tilde{\ell}_R\tilde{\ell}_R)}
O^{''L}_{ij}\to O^{''R}_{ij},\quad
 f_{\ell i}^L\to f_{\ell i}^R,\quad
 f_{\ell j}^L\to f_{\ell j}^R.\nonumber\label{lr_24}
\end{eqnarray}
\item {\sf The contributions of $s^3(\tilde{\chi}^0_i)$ and 
$s^1(\tilde{\chi}^0_j)$ are:}

The expressions for $\Sigma_P^{31}(\tilde{\chi}^0_i\tilde{\chi}^0_j)$
are obtained by exchanging
\begin{equation}
\eta_i\leftrightarrow \eta_j,\quad 
m_i\leftrightarrow m_j,\quad E_i\leftrightarrow E_j
\end{equation}
in eqs.~(\ref{ss13_zz})--(\ref{ss13_ll}) and also in the 
corresponding contributions from $\tilde{e}_R$ exchange.
\item {\sf The contributions of $s^1(\tilde{\chi}^0_i)$ and 
$s^2(\tilde{\chi}^0_j)$ are:}
\begin{eqnarray}
\Sigma_P^{12}(\tilde{\chi}^0_i\tilde{\chi}^0_j,ZZ)&=&
-4 \frac{g^4}{\cos^4\Theta_W}
|\Delta^s(Z)|^2 (c_{R}(ZZ)+c_{L}(ZZ))E_b^2 E_i q \sin^2\Theta
\nonumber\\&& 
Re(O^{''L}_{ij}) Im(O^{''L}_{ij}),\label{ss12_zz}\\
\Sigma_P^{12}(\tilde{\chi}^0_i\tilde{\chi}^0_j,Z \tilde{e}_L)&=&
\frac{g^4}{\cos^2\Theta_W}c_{L}(Z \tilde{e}_L)
E_b^2 E_i q \sin^2\Theta\nonumber\\
& &Im\Big\{\Delta^s(Z)[
-f^{L}_{\ell i}f^{L*}_{\ell j}\Delta^{u*}(\tilde{e}_L) O^{''L}_{ij}
+f^{L*}_{\ell i}f^{L}_{\ell j}\Delta^{t*}(\tilde{e}_L) 
O^{''L*}_{ij}]\Big\},\label{ss12_zl}\\
\Sigma_P^{12}(\tilde{\chi}^0_i\tilde{\chi}^0_j,\tilde{e}_L
\tilde{e}_L)&=&
\frac{g^4}{2} c_{L}(\tilde{e}_L\tilde{e}_L) E_b^2 E_i q\sin^2\Theta
Im\Big\{(f^{L*}_{\ell i})^2 (f^{L}_{\ell j})^2 \Delta^u(\tilde{e}_L)
\Delta^{t*}(\tilde{e}_L)\Big\}\label{ss12_ll}.
\end{eqnarray}
$\Sigma_P^{12}(\tilde{\chi}^0_i\tilde{\chi}^0_j,Z \tilde{e}_R),
 \Sigma_P^{12}(\tilde{\chi}^0_i\tilde{\chi}^0_j,\tilde{e}_R
\tilde{e}_R)$:
To obtain these quantities
one has to exchange\\
\phantom{$\Sigma_P^{12}(\tilde{\chi}^0_i\tilde{\chi}^0_j,Z \tilde{e}_R),
 \Sigma_P^{12}(\tilde{\chi}^0_i\tilde{\chi}^0_j,\tilde{e}_R
\tilde{e}_R)$: } 
in eqs.~(\ref{ss12_zl}) and (\ref{ss12_ll})
\begin{eqnarray}
&&\phantom{\Sigma_P^{12}(\tilde{\chi}^0_i\tilde{\chi}^0_j,Z \tilde{e}_R),
 \Sigma_P^{12}(\tilde{\chi}^0_i\tilde{\chi}^0_j,\tilde{e}_R
\tilde{e}_R)}
\Delta^{t}(\tilde{e}_L)\to\Delta^{t}(\tilde{e}_R),\quad
\Delta^{u}(\tilde{e}_L)\to\Delta^{u}(\tilde{e}_R),
\label{lr_25}\nonumber\\
&&\phantom{\Sigma_P^{12}(\tilde{\chi}^0_i\tilde{\chi}^0_j,Z \tilde{e}_R),
 \Sigma_P^{12}(\tilde{\chi}^0_i\tilde{\chi}^0_j,\tilde{e}_R
\tilde{e}_R)}
c_L(Z \tilde{e}_L)\to c_R(Z \tilde{e}_R),\quad 
c_L(\tilde{e}_L \tilde{e}_L)\to c_R(\tilde{e}_R \tilde{e}_R),
\label{lr_26}\nonumber\\
&&\phantom{\Sigma_P^{12}(\tilde{\chi}^0_i\tilde{\chi}^0_j,Z \tilde{e}_R),
 \Sigma_P^{12}(\tilde{\chi}^0_i\tilde{\chi}^0_j,\tilde{e}_R
\tilde{e}_R)}
O^{''L}_{ij}\to O^{''R}_{ij},\quad
 f_{\ell i}^L\to f_{\ell i}^R,\quad
 f_{\ell j}^L\to f_{\ell j}^R,\nonumber\label{lr_27}
\end{eqnarray} 
\phantom{$\Sigma_P^{12}(\tilde{\chi}^0_i\tilde{\chi}^0_j,Z \tilde{e}_R),
 \Sigma_P^{12}(\tilde{\chi}^0_i\tilde{\chi}^0_j,\tilde{e}_R
\tilde{e}_R)$ }
and to change the overall sign of 
eqs.~(\ref{ss12_zl}), (\ref{ss12_ll}).
\item {\sf The contributions of $s^2(\tilde{\chi}^0_i)$ and 
$s^1(\tilde{\chi}^0_j)$ are:}

The expressions for $\Sigma_P^{21}(\tilde{\chi}^0_i\tilde{\chi}^0_j)$
are obtained by exchanging
\begin{equation}
\eta_i\leftrightarrow \eta_j,\quad 
m_i\leftrightarrow m_j,\quad E_i\leftrightarrow E_j
\end{equation}
in eqs.~(\ref{ss12_zz})--(\ref{ss12_ll}) and in the 
corresponding contributions from $\tilde{e}_R$ exchange.
In addition, one also has to change the overall sign.
\item {\sf The contributions of $s^2(\tilde{\chi}^0_i)$ and 
$s^3(\tilde{\chi}^0_j)$ are:}
\begin{eqnarray}
\Sigma_P^{23}(\tilde{\chi}^0_i\tilde{\chi}^0_j,ZZ)&=&
 -4 \frac{g^4}{\cos^4\Theta_W}
|\Delta^s(Z)|^2 [c_{R}(ZZ)+c_{L}(ZZ)]\nonumber\\
& &  m_j E_b^2 q 
\sin\Theta \cos\Theta Re(O^{''L}_{ij}) Im(O^{''L}_{ij}),  
\label{ss23_zz}\\
\Sigma_P^{23}(\tilde{\chi}^0_i\tilde{\chi}^0_j,Z \tilde{e}_L)&=&
\frac{g^4}{\cos^2\Theta_W}c_{L}(Z \tilde{e}_L)
E_b^2 m_j q \sin\Theta\cos\Theta\nonumber\\&&
Im\Big\{\Delta^s(Z)[-f^{L}_{\ell i}f^{L*}_{\ell j}
\Delta^{u*}(\tilde{e}_L) O^{''L}_{ij}
+f^{L*}_{\ell i}f^{L}_{\ell j}\Delta^{t*}(\tilde{e}_L) 
O^{''L*}_{ij}]\Big\},\label{ss23_zl}\\ 
\Sigma_P^{23}(\tilde{\chi}^0_i\tilde{\chi}^0_j,
\tilde{e}_L \tilde{e}_L)&=&
\frac{g^4}{2} c_{L}(\tilde{e}_L\tilde{e}_L) 
E_b^2 m_j q \sin\Theta \cos\Theta\nonumber\\&&
Im\Big\{(f^{L*}_{\ell i})^2 (f^{L}_{\ell j})^2 \Delta^u(\tilde{e}_L)
\Delta^{t*}(\tilde{e}_L)\Big\}\label{ss23_ll}.
\end{eqnarray}
$\Sigma_P^{23}(\tilde{\chi}^0_i\tilde{\chi}^0_j,Z \tilde{e}_R),
 \Sigma_P^{23}(\tilde{\chi}^0_i\tilde{\chi}^0_j,
\tilde{e}_R \tilde{e}_R)$:
To obtain these quantities
one has to exchange\\
\phantom{$\Sigma_P^{23}(\tilde{\chi}^0_i\tilde{\chi}^0_j,Z \tilde{e}_R),
 \Sigma_P^{23}(\tilde{\chi}^0_i\tilde{\chi}^0_j,
\tilde{e}_R \tilde{e}_R)$: } 
 in eqs.~(\ref{ss23_zl}) and (\ref{ss23_ll})
\begin{eqnarray}
&&\phantom{\Sigma_P^{23}(\tilde{\chi}^0_i\tilde{\chi}^0_j,Z \tilde{e}_R),
 \Sigma_P^{23}(\tilde{\chi}^0_i\tilde{\chi}^0_j,\tilde{e}_R \tilde{e}_R)} 
\Delta^{t}(\tilde{e}_L)\to\Delta^{t}(\tilde{e}_R),\quad
\Delta^{u}(\tilde{e}_L)\to\Delta^{u}(\tilde{e}_R),
\label{lr_28}\nonumber\\
&&\phantom{\Sigma_P^{23}(\tilde{\chi}^0_i\tilde{\chi}^0_j,Z \tilde{e}_R),
 \Sigma_P^{23}(\tilde{\chi}^0_i\tilde{\chi}^0_j,
\tilde{e}_R \tilde{e}_R)} 
c_L(Z \tilde{e}_L)\to c_R(Z \tilde{e}_R),\quad 
c_L(\tilde{e}_L \tilde{e}_L)\to c_R(\tilde{e}_R \tilde{e}_R),
\label{lr_29}\nonumber\\
&&\phantom{\Sigma_P^{23}(\tilde{\chi}^0_i\tilde{\chi}^0_j,Z \tilde{e}_R),
 \Sigma_P^{23}(\tilde{\chi}^0_i\tilde{\chi}^0_j,
\tilde{e}_R \tilde{e}_R)} 
O^{''L}_{ij}\to O^{''R}_{ij},\quad
 f_{\ell i}^L\to f_{\ell i}^R,\quad
 f_{\ell j}^L\to f_{\ell j}^R,\label{lr_30}\nonumber
\end{eqnarray}
\phantom{$\Sigma_P^{23}(\tilde{\chi}^0_i\tilde{\chi}^0_j,Z \tilde{e}_R),
 \Sigma_P^{23}(\tilde{\chi}^0_i\tilde{\chi}^0_j,\tilde{e}_R \tilde{e}_R)$ }  
and to change the overall sign of eqs.~(\ref{ss23_zl}), (\ref{ss23_ll}).
\item {\sf The contributions of $s^3(\tilde{\chi}^0_i)$ and 
$s^2(\tilde{\chi}^0_j)$ are:}

The expressions for $\Sigma_P^{32}(\tilde{\chi}^0_i\tilde{\chi}^0_j)$
are obtained by exchanging
\begin{equation}
m_i\leftrightarrow m_j,\quad E_i\leftrightarrow E_j
\end{equation}
in eqs.~(\ref{ss23_zz})--(\ref{ss23_ll}) and in the 
corresponding contributions from $\tilde{e}_R$ exchange.
In addition, one also has to change the overall sign.
\end{enumerate}
Note that
\begin{itemize}
\item all contributions of transverse polarizations 
$s^1(\tilde{\chi}^0_i)$, $s^2(\tilde{\chi}^0_i)$, 
 $s^1(\tilde{\chi}^0_j)$, $s^2(\tilde{\chi}^0_j)$
vanish in forward and backward direction;
\item at threshold all spin-spin terms of transverse polarizations
$s^2(\tilde{\chi}^0_i)$, $s^2(\tilde{\chi}^0_j)$
vanish proportional to the momentum of the neutralinos.
\end{itemize}
\section{Decay matrix}
In the following we give the analytical formulae for the
decay matrices 
$D(\tilde{\chi}^0_i)$, $\Sigma_D^a(\tilde{\chi}^0_i)$ 
for the decay $\tilde{\chi}_i^{0}(p_3)\to\tilde{\chi}_k^0(p_5) 
+\ell^{+}(p_6)+\ell^{-}(p_7)$,
and $D(\tilde{\chi}^0_j)$, $\Sigma_D^b(\tilde{\chi}^0_j)$ for the decay
$\tilde{\chi}_j^{0}(p_4)\to\tilde{\chi}_l^0(p_8)
+\ell^{+}(p_9)+\ell^{-}(p_{10})$. 
We present them in covariant form. They have to be inserted in 
eq.~({\ref{eq_14}) to obtain the amplitude squared for the combined
  process of neutralino production and decay.
\subsection{Neutralino polarization independent quantities}
The expression $D(\tilde{\chi}^0_i)$ of eq.~(\ref{eq_14}) which is  
independent of the polarization vector $s^a(\tilde{\chi}^0_i)$
has the following decomposition:
\begin{equation}
D(\tilde{\chi}^0_i)=D(\tilde{\chi}^0_i,Z Z)
+D(\tilde{\chi}^0_i,Z \tilde{\ell}_L)+D(\tilde{\chi}^0_i,Z
\tilde{\ell}_R)
+D(\tilde{\chi}^0_i,\tilde{\ell}_L \tilde{\ell}_L)
+D(\tilde{\chi}^0_i,\tilde{\ell}_R \tilde{\ell}_R).\label{eq_30}
\end{equation}
The analytical expressions for $D(\tilde{\chi}^0_i)$,
eq.~(\ref{eq_30}), read: 
\begin{eqnarray}
D(\tilde{\chi}^0_i,Z Z)&=& 8 \frac{g^4}{\cos^4\Theta_W} |\Delta^{s_i}(Z)|^2 
\nonumber\\&&
(L_{\ell}^2+R_{\ell}^2)
\Big[ |O^{''L}_{ki}|^2 (g_1+g_2)
+[(Re O^{''L}_{ki})^2 -(Im O^{''L}_{ki})^2] g_3  \Big], \label{d_zz}\\
D(\tilde{\chi}^0_i,Z \tilde{\ell}_L)&=&4 
\frac{g^4}{\cos^2\Theta_W} L_{\ell}
Re\Big\{\Delta^{s_i}(Z)
\big[f^L_{\ell i} f^{L*}_{\ell k}
\Delta^{t_i*}(\tilde{\ell}_L)
(2O^{''L}_{ki} g_1 +O^{''L*}_{ki} g_3) \nonumber\\
& & \phantom{4 
\frac{g^4}{\cos^2\Theta_W} L_{\ell}
Re\Big\{\Delta_i(Z)}
+f^{L*}_{\ell i} f^{L}_{\ell k}
\Delta^{u_i*}(\tilde{\ell}_L)
(2O^{''L*}_{ki} g_2 +O^{''L}_{ki} g_3) 
\big]\Big\},\label{d_zl}\\
D(\tilde{\chi}^0_i,\tilde{\ell}_L \tilde{\ell}_L)&=&
2 g^4 \Big[ 
|f^{L}_{\ell i}|^2 |f^L_{\ell k}|^2
\big(|\Delta^{t_i}(\tilde{\ell}_L)|^2 g_1
+|\Delta^{u_i}(\tilde{\ell}_L)|^2 g_2\big)\nonumber\\
& &+Re\big\{(f^{L*}_{\ell i})^2 (f^L_{\ell k})^2
\Delta^{t_i}(\tilde{\ell}_L)
\Delta^{u_i*}(\tilde{\ell}_L)\big\} g_3 \Big],\label{d_ll}
\end{eqnarray}
where we have introduced the following combinations of 
scalar products:
\begin{eqnarray}
g_1&=&(p_5 p_{7}) (p_3 p_6), \label{eq_31}\\
g_2&=&(p_5 p_6) (p_3 p_{7}), \label{eq_32}\\
g_3&=&(\eta_i \eta_k m_i m_k) (p_6 p_{7})\label{eq_33}.
\end{eqnarray}
The propagators 
are denoted by $\Delta^{s_i}(Z)$,
 $\Delta^{t_i}(\tilde{\ell}_{L,R})$, $\Delta^{u_i}(\tilde{\ell}_{L,R})$ and 
are defined analogously to eq.~(\ref{eq_11}), with $s_i$, $t_i$, $u_i$ as 
defined after eq.~(\ref{eq_10}).\\

$D(\tilde{\chi}^0_i,Z \tilde{\ell}_R),
 D(\tilde{\chi}^0_i,\tilde{\ell}_R \tilde{\ell}_R):$
To obtain these quantities
one has to exchange in eqs.~(\ref{d_zl})\\
\phantom{$D(\tilde{\chi}^0_i, Z \tilde{\ell}_R),
 D(\tilde{\chi}^0_i,\tilde{\ell}_R \tilde{\ell}_R): $ \quad}
 and (\ref{d_ll})
\begin{eqnarray}
&&\phantom{D(\tilde{\chi}^0_i,Z \tilde{\ell}_R),
 D(\tilde{\chi}^0_i,\tilde{\ell}_R \tilde{\ell}_R): }
 \Delta^{t_i}(\tilde{\ell}_L)\to\Delta^{t_i}(\tilde{\ell}_R),\quad
 \Delta^{u_i}(\tilde{\ell}_L)\to\Delta^{u_i}(\tilde{\ell}_R),
\label{lr_31}\nonumber\\
&&\phantom{D(\tilde{\chi}^0_i,Z \tilde{\ell}_R),
 D(\tilde{\chi}^0_i,\tilde{\ell}_R \tilde{\ell}_R): }
 O^{''L}_{ki}\to O^{''R}_{ki},\quad
 f_{\ell i}^L\to f_{\ell i}^R,\quad L_{\ell} \to R_{\ell}.
\label{lr_32}\nonumber
\end{eqnarray}
The expressions $D_j(\tilde{\chi}^0_j)$,
 eq.~(\ref{eq_14}), for the decay  
$\tilde{\chi}^{0}_j(p_4)\to 
\tilde{\chi}^0_l(p_8)+\ell^{+}(p_9)+\ell^{-}(p_{10})$ 
and the corresponding scalar products 
 are obtained by the following substitutions in 
eqs.~(\ref{d_zz})--(\ref{eq_33}):
\begin{eqnarray}
& & p_5 \to p_8, p_6 \to p_9, p_{7} \to p_{10}, \quad
m_i \to m_j, m_k \to m_l, \quad 
\eta_i\to\eta_j, \eta_k\to \eta_l, \label{eq_40}\\
& & O^L_{ki}\to O^{L}_{lj},\quad 
O^R_{ki}\to O^{R}_{lj},\label{eq_41}\\
& & \Delta^{s_i}(Z)\to \Delta^{s_j}(Z), 
\quad \Delta^{t_i}(\tilde{\ell}_{L,R})
\to \Delta^{t_j}(\tilde{\ell}_{L,R}),\quad 
\Delta^{u_i}(\tilde{\ell}_{L,R})
\to \Delta^{u_j}(\tilde{\ell}_{L,R})
\label{eq_42}.
\end{eqnarray}
\subsection{Neutralino polarization dependent quantities}
We first give $\Sigma_D^a(\tilde{\chi}^0_i)$ 
of eq.~(\ref{eq_14}) which contains 
the polarization vector $s^a(\tilde{\chi}^{0}_i)$: 
\begin{equation}
\Sigma_D^a(\tilde{\chi}^0_i)=
\Sigma_D^a(\tilde{\chi}^0_i,ZZ)
+\Sigma_D^a(\tilde{\chi}^0_i,Z \tilde{\ell}_L)
+\Sigma_D^a(\tilde{\chi}^0_i,Z \tilde{\ell}_R)
+\Sigma_D^a(\tilde{\chi}^0_i,\tilde{\ell}_L \tilde{\ell}_L)
+\Sigma_D^a(\tilde{\chi}^0_i,\tilde{\ell}_R \tilde{\ell}_R).\label{eq_36}
\end{equation} 
The analytical expressions for $\Sigma^a_D(\tilde{\chi}^0_i)$,
eq.~(\ref{eq_36}), read:
\begin{eqnarray}
\Sigma_D^a(\tilde{\chi}^0_i,ZZ)&=&
8 \frac{g^4}{\cos^4\Theta_W} |\Delta^{s_i}(Z)|^2 
(R^2_{\ell}-L_{\ell}^2) \nonumber\\
& &\Big[-[(Re O^{''L}_{ki})^2 -(Im O^{''L}_{ki})^2]g^a_3+
|O^{''L}_{ki}|^2(g^a_1-g^a_2)\Big],\label{ds_zz}\\ 
\Sigma_D^a(\tilde{\chi}^0_i,Z \tilde{\ell}_L)&=& 
\frac{4 g^4}{\cos^2\Theta_W}L_{\ell}
 Re\Big\{\Delta^{s_i}(Z)
\Big[
f^L_{\ell i} f^{L*}_{\ell k}
\Delta^{t_i*}(\tilde{\ell}_L)
\big(-2 O^{''L}_{ki} g^a_1 +O^{''L*}_{ki} (g^a_3-g^a_4)\big)\nonumber\\
& &\phantom{ 
\frac{4 g^4}{\cos^2\Theta_W}L_{\ell}
 Re\Big\{\Delta_i(Z}
+f^{L*}_{\ell i} f^{L}_{\ell k}
\Delta^{u_i*}(\tilde{\ell}_L)
\big(2 O^{''L*}_{ki} g^a_2 +O^{''L}_{ki} (g^a_3-g^a_4)\big)
\Big]\Big\},\label{ds_zl}\\
\Sigma_D^a(\tilde{\chi}^0_i,\tilde{\ell}_L \tilde{\ell}_L)&=&
2 g^4 \Big[
|f^{L}_{\ell i}|^2 |f^L_{\ell k}|^2
[|\Delta^{u_i}(\tilde{\ell}_L)|^2 g_2^a 
-|\Delta^{t_i}(\tilde{\ell}_L)|^2 g_1^a]\nonumber\\
& & \phantom{2 g^4 \Big[} 
+Re\big\{ (f^{L*}_{\ell i})^2 (f^L_{\ell k})^2 
\Delta^{t_i}(\tilde{\ell}_L)
\Delta^{u_i*}(\tilde{\ell}_L)(g_3^a+g_4^a)\big\}\Big],\label{ds_ll}
\end{eqnarray}
where we have introduced the following abbrevations involving
the polarization vector $s^a(\tilde{\chi}^0_i)$,
eqs.~(\ref{eq_21})--(\ref{eq_23}), with $a=$1, 2, 3:
\begin{eqnarray}
g^a_1&=&\eta_i m_i (p_5 p_{7}) (p_6 s^a(\tilde{\chi}^0_i)), \label{eq_38}\\
g^a_2&=&\eta_i m_i (p_5 p_6) (p_{7} s^a(\tilde{\chi}^0_i)), \\
g^a_3&=&\eta_k m_k [(p_3 p_6) (p_{7} s^a(\tilde{\chi}^0_i))
            -(p_3 p_{7}) (p_6 s^a(\tilde{\chi}^0_i))],\label{eq_39}\\
g^a_4&=&i \eta_k m_k \epsilon_{\mu \nu \rho \tau}s^{a \mu}(\tilde{\chi}^0_i)
 p_3^{\nu} p_{7}^{\rho} p_6^{\tau}.\label{eq_37}
\end{eqnarray}

$\Sigma_D^a(\tilde{\chi}^0_i,Z \tilde{\ell}_R)$,
 $\Sigma_D^a(\tilde{\chi}^0_i,\tilde{\ell}_R \tilde{\ell}_R)$:
To obtain these quantities
one has to exchange in ~eqs.(\ref{ds_zl})\\
\phantom{$\Sigma_D^a(\tilde{\chi}^0_i,Z \tilde{\ell}_R),
\Sigma_D^a(\tilde{\chi}^0_i,\tilde{\ell}_R \tilde{\ell}_R)$: \quad}
 and (\ref{ds_ll})
\begin{eqnarray}
&& \phantom{\Sigma_D^a(\tilde{\chi}^0_i,Z \tilde{\ell}_R),
\Sigma_D^a(\tilde{\chi}^0_i,\tilde{\ell}_R \tilde{\ell}_R): }
 \Delta^{t_i}(\tilde{\ell}_L)\to \Delta^{t_i}(\tilde{\ell}_R),\quad
 \Delta^{u_i}(\tilde{\ell}_L)\to \Delta^{u_i}(\tilde{\ell}_R),
\nonumber
\label{lr_33}\\
&& \phantom{\Sigma_D^a(\tilde{\chi}^0_i,Z \tilde{\ell}_R),
\Sigma_D^a(\tilde{\chi}^0_i,\tilde{\ell}_R \tilde{\ell}_R): }
O^{''L}_{ki}\to O^{''R}_{ki},\quad
 f_{\ell i}^L\to f_{\ell i}^R,\quad L_{\ell} \to R_{\ell}.\nonumber\label{lr_34}
\end{eqnarray}
\phantom{$\Sigma_D^a(\tilde{\chi}^0_i,Z \tilde{\ell}_R),
\Sigma_D^a(\tilde{\chi}^0_i,\tilde{\ell}_R \tilde{\ell}_R)$: \quad}
 In addition, one has to change the  sign of $g_1^a$, $g_2^a$, $g_3^a$,
but\\
\phantom{$\Sigma_D^a(\tilde{\chi}^0_i,Z \tilde{\ell}_R),
\Sigma_D^a(\tilde{\chi}^0_i,\tilde{\ell}_R \tilde{\ell}_R)$: \quad}
 not of $g_4^a$. 

\noindent The expression $g_4^a$ can be expanded in triple product 
correlations which are sensitive to the component of the spin vector
perpendicular to the scattering plane. 

The corresponding expressions $\Sigma_D^b(\tilde{\chi}^0_j)$, 
eq.~(\ref{eq_14}), for the 
decay 
$\tilde{\chi}^{0}_j(p_4)\to 
\tilde{\chi}^0_l(p_8)+\ell^{+}(p_9)+\ell^{-}(p_{10})$ are obtained by  
the same substitutions as eqs.~(\ref{eq_40})--(\ref{eq_42}), 
 and the additional substitution $s^a(\tilde{\chi}^0_i) \to
 s^b(\tilde{\chi}^0_j)$ in eqs.~(\ref{eq_38})--(\ref{eq_37}). 
\section{Numerical Results}
\vspace{-.3cm}
In the following numerical analysis we study $e^{+}e^{-}\to
\tilde{\chi}^0_1 \tilde{\chi}^0_2$ with $\tilde{\chi}^0_2\to
\tilde{\chi}^0_1 e^{+}e^{-}$ for various polarizations of the
$e^{-}$ beam.
The calculations are done in the MSSM.
We take the parameters $M'$,
$M$, $\mu,$ $\tan\beta$  real. Since we want to study the influence
of the parameter $M'$ we do not use a relation between $M'$ and
$M$. We will also study the dependence on the selectron masses
$m_{\tilde{e}_{L}}$, $m_{\tilde{e}_{R}}$. 

We shall choose three different examples of parameter sets. In all these
examples we choose $M=152$~GeV, $\mu=316$~GeV, $\tan\beta=3$, and vary
$M'$ between 40 GeV and 160 GeV. Especially the mass of the
$\tilde{\chi}^0_1$ is very sensitive to $M'$. 
For these parameters $\tilde{\chi}^0_1$ and 
$\tilde{\chi}^0_2$ are dominantly gauginos and have small couplings to $Z^0$.
For the selectron masses we take
\begin{itemize}
\item[i)] $m_{\tilde{e}_L}=1000$~GeV, $m_{\tilde{e}_R}=200$~GeV; 
\item[ii)] $m_{\tilde{e}_L}=200$~GeV, $m_{\tilde{e}_R}=1000$~GeV; 
\item[iii)] $m_{\tilde{e}_L}=176$~GeV, $m_{\tilde{e}_R}=161$~GeV. 
\end{itemize}

In i) and ii) we want to study the influence of $\tilde{e}_L$ and 
$\tilde{e}_R$ exchange for large slepton mass splitting. 
Scenario ii) with 
$m_{\tilde{e}_R}>m_{\tilde{e}_L}$ may be realized in  extended
SUSY models \cite{hesselbach}.
For $M'=78.7$~GeV, example iii) corresponds to the mSUGRA scenario
studied in \cite{blair}. 
In i) and ii) we want to study the influence of $\tilde{e}_L$ and 
$\tilde{e}_R$ exchange for large slepton mass splitting. 

We present results for the cross section
\begin{equation}
\sigma_e=\sigma(e^{+}e^{-}\to\tilde{\chi}^0_1\tilde{\chi}^0_2)
BR(\tilde{\chi}^0_2\to\tilde{\chi}^0_1 e^{+} e^{-}),\label{sbr}
\end{equation}
and the
forward--backward asymmetry 
\begin{equation} 
A_{FB}=\frac{\sigma_e(\cos\Theta_{-}>0)-\sigma_e(\cos\Theta_{-}<0)}
{\sigma_e(\cos\Theta_{-}>0)+\sigma_e(\cos\Theta_{-}<0)}
\label{afb}\end{equation}
of the electron
from the decay 
$\tilde{\chi}^{0}_2 \to \tilde{\chi}^0_1 e^{+} e^{-}$.
In eq.~(\ref{afb})
$\Theta_{-}$ is the angle between the incoming electron beam and the
outgoing $e^{-}$.

The forward--backward asymmetry $A_{FB}$ is largest near the production 
threshold. We therefore study in all three examples $\sigma_e$
and $A_{FB}$ at
$\sqrt{s}=m_{\tilde{\chi}^0_1}+m_{\tilde{\chi}^0_2}+50$~GeV, and 
in example iii) 
also at $\sqrt{s}=500$~GeV. As for the polarization of the $e^{-}$
beam we take $P^3_{-}=\pm90\%$.
\subsection{Cross sections}
We first study the $M'$ dependence of 
$\sigma_e=\sigma(e^{+}e^{-}\to \tilde{\chi}^0_1\tilde{\chi}^0_2)
 BR(\tilde{\chi}^0_2\to \tilde{\chi}^0_1 e^{+} e^{-})$ 
near the production threshold 
($\sqrt{s}=m_{\tilde{\chi}^0_1}+m_{\tilde{\chi}^0_2}+50$~GeV). 
We begin with case i), where $\tilde{e}_L$ exchange is suppressed.
Fig.~\ref{sitt50_12} shows the corresponding $M'$ dependence for unpolarized
beams and for the $e^{-}$ beam polarizations $P_{-}^3=+90\%$ and 
$P_{-}^3=-90\%$, with $M, \mu$ and $\tan\beta$ as given above.
Clearly, a right polarized $e^{-}$ beam yields the largest cross
section because it enhances the $\tilde{e}_R$ exchange. 
The production cross section for 
$e^{+}e^{-}\to \tilde{\chi}^0_1 \tilde{\chi}^0_2$ has a maximum at
$M'\approx 130$~GeV, where also the $\tilde{e}_R$ exchange
contribution is maximal.
The cross section $\sigma_e$, eq. (\ref{sbr}), has its maximum 
at $M'\approx 118$~GeV.
This shift is due to the fact that the leptonic decay branching ratio of 
$\tilde{\chi}^0_2$ has a maximum at $M'\approx 118$~GeV and then
strongly decreases.

Obviously, the characters of $\tilde{\chi}^0_1$ and $\tilde{\chi}^0_2$ change
with varying $M'$. With increasing $M'$ the $\tilde{B}$ component of 
$\tilde{\chi}^0_1$ decreases and the $W^3$-ino and the higgsino components
increase. The opposite is true for $\tilde{\chi}^0_2$. 
The $Z^0$ couplings are small and almost constant up  to 
$M'\approx 120$~GeV, $O^{''L}_{12}\approx 0.015$, and
decrease for larger $M'$. The product of the $\tilde{e}_R$
couplings, $|f_{e 1}^R f_{e 2}^R|$, has a
maximum at $M'\approx 130$~GeV.

We compare this with case ii), where $\tilde{e}_R$ exchange is 
suppressed. Fig.~\ref{sitt50_21} shows the corresponding $M'$ dependence. 
Now a left polarized $e^{-}$ beam leads to the largest cross section because 
the $\tilde{e}_L$ exchange is favoured. There is a maximum at 
$M'\approx 60$~GeV and a minimum at $M'\approx 120$~GeV. 
The maximum at $M'\approx 60$~GeV can be explained by a corresponding
maximum of the leptonic branching ratio. The minimum at $M'\approx 120$~GeV 
is due to the vanishing of $e\tilde{e}_L \tilde{\chi}^0_1$ coupling
$f^L_{e 1}$ at this value of $M'$.

In example iii) the mass difference between $\tilde{e}_L$ and 
$\tilde{e}_R$ is small. Therefore, $\tilde{e}_L$ and $\tilde{e}_R$ 
exchange contribute. We show in Figs.~\ref{sitt50_m0} and \ref{sit_m0}
the $M'$ dependence for this case near threshold and at $\sqrt{s}=500$~GeV, 
respectively. For right polarized $e^{-}$ beams the cross section 
behaviour is similar to that of case i), and for left polarized $e^{-}$ beams
it is similar to that of case ii). At $\sqrt{s}=500$~GeV the cross section is
about a factor 2 bigger than near threshold but has a similar $M'$ 
dependence.
In all cases there is a small step at about $M'=42-44$~GeV, which is
due to the opening of the two-body decay
$\tilde{\chi}^0_2\to\tilde{\chi}^0_1 Z^0$.
\subsection{Lepton forward--backward asymmetries}
In this subsection we study the $M'$ dependence of the forward--backward
asymmetry $A_{FB}$ of the decay electron $e^{-}$, as defined in 
eq.~(\ref{afb}). 
The decay electron angular distributions and the corresponding
forward--backward asymmetry are very sensitive to the spin correlations 
$\Sigma_P^a\Sigma_D^a$, ($\Sigma_P^b\Sigma_D^b$), eq.~(\ref{eq_14}), 
and are the
result of a complex interplay between production and decay.
As the spin correlations between production and decay are strongest near
threshold, the forward--backward asymmetry  will also be largest there.

We show in Figs.~\ref{asyt50_12}, \ref{asyt50_21}, and \ref{asyt50_m0} 
$A_{FB}$ near threshold as a function of $M'$ for
the cases i), ii), and iii), respectively. 
As can be seen $A_{FB}$ is very sensitive to the masses of
$\tilde{e}_L$ and $\tilde{e}_R$, and the mass splitting between them.
In all cases $A_{FB}$ has a pronounced $M'$ dependence. The selectron
couplings $f_{e i}^L$ and $f_{e i}^R$, $i=$1, 2, exhibit a
characteristic $M'$ dependence, which is reflected in the $M'$
behaviour of $A_{FB}$. Moreover, by choosing different $e^{-}$ beam
polarizations the $\tilde{e}_L$ and $\tilde{e}_R$ contributions can be 
enhanced or suppressed.

The small dip of the asymmetry at $M'=42-44$~GeV is due to the opening
of the two-body decay $\tilde{\chi}^0_2\to\tilde{\chi}^0_1 Z^0$.

The behaviour of $A_{FB}$ in Fig.~\ref{asyt50_21} at about $M'=115-125$~GeV
is due to the vanishing of $e \tilde{e}_L \tilde{\chi}^0_1$ coupling
$f^L_{e 1}$ at $M'\approx 120$~GeV and a complicated interplay
between the $Z^0$, $\tilde{e}_L$ and $\tilde{e}_R$ contributions,
which are all very small (see Fig.~\ref{sitt50_21}).

In Fig.~\ref{asy_m0} we show the $M'$ dependence of $A_{FB}$ at 
$\sqrt{s}=500$~GeV for case iii). This is very similar to that near 
threshold, Fig.~\ref{asyt50_m0}, but the magnitude is smaller by a 
factor 2 to 3, because with increasing $\sqrt{s}$ the spin correlations 
decrease.

A numerical analysis for both beams polarized has been given in
\cite{cracow98} and will be continued in \cite{gudi_prep}.
\section{Summary}
We have given the full analytical expressions for the differential
cross section for $e^{+}e^{-}\to\tilde{\chi}^{0}_i \tilde{\chi}^{0}_j$
with polarized beams and the subsequent 
leptonic decays
$\tilde{\chi}^0_{i}\to\ell^{+}\ell^{-}\tilde{\chi}^0_{k}$ and 
$\tilde{\chi}^0_{j}\to\ell^{+}\ell^{-}\tilde{\chi}^0_{l}$,
taking into
account the complete spin correlations between production and decay.
The production spin density matrix is presented in the laboratory
system.
The formulae for the decay processes are written covariantly involving
explicitly the neutralino polarization vectors. 
When combining the production and decay process the polarization 
vectors in the laboratory system as given in 
eqs.~(\ref{eq_21})--(\ref{eq_26}) have to be taken.

We have presented numerical results for the cross section and the lepton
forward--backward asymmetry for 
$e^{+}e^{-}\to\tilde{\chi}^0_1\tilde{\chi}^0_2$, 
$\tilde{\chi}^0_2 \to\tilde{\chi}^0_1 e^{+} e^{-}$. We have studied the
dependence on the parameter $M'$ for various mass splittings between
$\tilde{e}_L$ and $\tilde{e}_R$ and different $e^{-}$ beam polarizations.

The cross section $\sigma_e$ 
shows a characteristic dependence on $M'$ and the masses of
the exchanged selectrons as well as on the beam polarization.

The lepton forward--backward asymmetry $A_{FB}$ can only be explained by the
presence of spin correlations between production and decay, as it would 
be zero in the production process alone. $A_{FB}$ depends very sensitively 
on the SUSY parameters and the beam polarizations. Therefore, this quantity is
an additional useful observable 
for a more precise determination of the parameters.
Different beam polarizations help disentangle the contribution from
$\tilde{e}_L$ and $\tilde{e}_R$ exchange. 
\section{Acknowledgement}
We are grateful to S.~Hesselbach for many useful discussions and
to V.~Latussek for his support in the development of the
numerical program. We are also grateful to W.~Porod and S.~Hesselbach 
for providing the computer programs for neutralino widths.
G.M.-P. was supported by {\it
  Friedrich-Ebert-Stiftung}. This work was also supported by 
the German Federal Ministry for
Research and Technology (BMBF) under contract number
05 7WZ91P (0), by the Deutsche Forschungsgemeinschaft under
contract Fr 1064/2-2, and the `Fonds zur
F\"orderung der wissenschaftlichen Forschung' of Austria, Projects
No. P10843-PHY and P13139-PHY. 

\vspace*{.6cm}
\begin{appendix}
\noindent{\Large\bf Appendices}
\setcounter{equation}{0}
\renewcommand{\thesubsection}{\Alph{section}.\arabic{subsection}}
\renewcommand{\theequation}{\Alph{section}.\arabic{equation}}
\vspace*{-.4cm}
\section{Amplitudes}
We give the helicity amplitudes $T_P^{\lambda_i\lambda_j}(\alpha)$ for
production and $T_{D,\lambda_i}(\alpha)$, $T_{D,\lambda_j}(\alpha)$ for
the decays, 
corresponding to the Feynman diagrams in Fig.~1
($\alpha$ denotes the channel). 

The amplitudes $T_P^{\lambda_i\lambda_j}(\alpha)$ for the production,
$e^{-}(p_1) e^{+}(p_2)\to \tilde{\chi}^0_i(p_3)\tilde{\chi}^0_j(p_4)$ read:
\begin{eqnarray}
& &\mbox{\hspace*{-.5cm}}
T_P^{\lambda_i\lambda_j}(s,Z)
=\frac{g^2}{\cos^2\Theta_W}
 \Delta^s(Z)\bar{v}(p_2)\gamma^{\mu}(L_{\ell} P_L+R_{\ell}
P_R)u(p_1)\nonumber\\ 
& &\mbox{\hspace*{-.5cm}}
\phantom{T_P^{\lambda_i\lambda_j}(s)=\frac{g^2}{\cos^2\Theta_W}\Delta(Z)}
\bar{u}(p_4,\lambda_j)\gamma_{\mu} 
(O^{''L}_{ji} P_L+O^{''R}_{ji} P_R)v(p_3,\lambda_i),\label{eq_a1}\\
& &\mbox{\hspace*{-.5cm}}
T_P^{\lambda_i\lambda_j}(t,\tilde{\ell}_{L})
=-g^2 f^L_{\ell i}f^{L*}_{\ell j}
\Delta^t(\tilde{\ell}_L)\bar{v}(p_2) P_R v(p_3,\lambda_i)
\bar{u}(p_4,\lambda_j)P_L u(p_1),\label{eq_a2a}\\
& &\mbox{\hspace*{-.5cm}}
T_P^{\lambda_i\lambda_j}(t,\tilde{\ell}_{R})
=-g^2 f^R_{\ell i}f^{R*}_{\ell j}
\Delta^t(\tilde{\ell}_R)\bar{v}(p_2) P_L v(p_3,\lambda_i)
\bar{u}(p_4,\lambda_j)P_R u(p_1),\label{eq_a2}\\
& &\mbox{\hspace*{-.5cm}}
T_P^{\lambda_i\lambda_j}(u,\tilde{\ell}_L)
=g^2 f^{L*}_{\ell i} f^L_{\ell j}
\Delta^u(\tilde{\ell}_L)\bar{v}(p_2) P_R v(p_4,\lambda_j)
\bar{u}(p_3,\lambda_i)P_L u(p_1),\label{eq_a3a}\\
& &\mbox{\hspace*{-.5cm}}
T_P^{\lambda_i\lambda_j}(u,\tilde{\ell}_R)
=g^2 f^{R*}_{\ell i} f^R_{\ell j}
\Delta^u(\tilde{\ell}_R)\bar{v}(p_2) P_L v(p_4,\lambda_j)
\bar{u}(p_3,\lambda_i)P_R u(p_1).\label{eq_a3}\\
\end{eqnarray}
The amplitudes  $T_{D,\lambda_i}(\alpha)$ for the decay
of the 
$\tilde{\chi}^0_i(p_3)\to\tilde{\chi}^0_k(p_5)\ell^{+}(p_6)\ell^{-}(p_7)$ read:
\begin{eqnarray}
& &\mbox{\hspace*{-.5cm}}
T_{D, \lambda_i}(s_i)
=-\frac{g^2}{\cos^2\Theta_W}
\Delta^{s_i}(Z) \bar{u}(p_7) \gamma^{\mu} (L_{\ell} P_L+R_{\ell} P_R) v(p_6)
\nonumber\\
& &\mbox{\hspace*{-.5cm}}
\phantom{T_{D, \lambda_i}(s_i)=-\frac{g^2}{\cos^2\Theta_W}\Delta^{s_i}(Z)}
\bar{u}(p_5) \gamma_{\mu} (O^{''L}_{ki} P_L+O^{''R}_{ki} P_R)
 u(p_3, \lambda_i), \label{eq_a4}\\
& &\mbox{\hspace*{-.5cm}}
T_{D, \lambda_i}(t_i,\tilde{\ell}_L)
=- g^2 f^L_{\ell k} f^{L*}_{\ell i}
\Delta^{t_i}(\tilde{\ell}_L)\bar{u}(p_7) P_R v(p_5)
\bar{v}(p_3, \lambda_i) P_L v(p_6),\label{eq_a5a}\\
& &\mbox{\hspace*{-.5cm}}
T_{D, \lambda_i}(t_i,\tilde{\ell}_R)
=-g^2 f^R_{\ell k} f^{R*}_{\ell i}
\Delta^{t_i}(\tilde{\ell}_R)\bar{u}(p_7) P_L v(p_5)
\bar{v}(p_3, \lambda_i) P_R v(p_6), \label{eq_a5}\\
& &\mbox{\hspace*{-.5cm}}
T_{D, \lambda_i}(u_i,\tilde{\ell}_L)
=+g^2 f^L_{\ell i} f^{L*}_{\ell k}
\Delta^{u_i}(\tilde{\ell}_L)\bar{u}(p_7) P_R u(p_3, \lambda_i)
\bar{u}(p_5) P_L v(p_6),\label{eq_a6a}\\
& &\mbox{\hspace*{-.5cm}}
T_{D, \lambda_i}(u_i,\tilde{\ell}_R)
=+g^2 f^R_{\ell i} f^{R*}_{\ell k}
\Delta^{u_i}(\tilde{\ell}_R)\bar{u}(p_7) P_L u(p_3, \lambda_i)
\bar{u}(p_5) P_R v(p_6).\label{eq_a6}
\end{eqnarray}
The corresponding amplitudes for the decay $T_{D,\lambda_j}(\alpha)$ 
of the
$\tilde{\chi}^0_j(p_4)\to\tilde{\chi}^0_l(p_8)
\ell^{+}(p_9)\ell^{-}(p_{10})$ 
are obtained by exchanging in eqs.~(\ref{eq_a4})--(\ref{eq_a6}):
\begin{eqnarray}
& & s_i\to s_j, t_i\to t_j, u_i\to u_j, \quad
\Delta^{s_i}\to\Delta^{s_j}, \Delta^{t_i}\to\Delta^{t_j},
\Delta^{u_i}\to\Delta^{u_j},\label{eq_a7a}\\
& & p_5 \to p_8, p_6 \to p_9, p_{7} \to p_{10}, \quad
O^L_{ki}\to O^{L}_{lj}, O^R_{ki}\to O^{R}_{lj}.\label{eq_a7}
\end{eqnarray}
\section{Spin Formalism}
\setcounter{equation}{0}
 The amplitude for the whole process, eq.~(\ref{eq_12}), is
\begin{equation}
T=\Delta(\tilde{\chi}^{+}_i) \Delta(\tilde{\chi}^{-}_j)
\sum_{\lambda_i, \lambda
_j}T_P^{\lambda_i \lambda_j} 
T_{D, \lambda_i}T_{D, \lambda_j},\label{B_1}
\end{equation}
 with the helicity amplitude $T_P^{\lambda_i \lambda_j}$ 
for the production process and $T_{D,\lambda_i}$, $T_{D,\lambda_j}$
 for the decay processes, 
 and the propagators $\Delta(\tilde{\chi}^{\pm}_{i,j})=1/[p_{i,j}-m_{i,j}^2+i
m_{i,j}\Gamma_{i,j}]$.
Here $\lambda_{i,j}, p_{i,j}$, $m_{i,j}$, $\Gamma_{i,j}$ 
denote the helicity, four--momentum squared, mass and width of
$\tilde{\chi}^{\pm}_{i,j}$. The amplitude squared
\begin{equation}
|T|^2=|\Delta(\tilde{\chi}^{+}_i)|^2 |\Delta(\tilde{\chi}^{-}_j)|^2 
\rho^{\lambda_i\lambda_j \lambda_i'\lambda_j'}_P
\rho_{D,\lambda_i'\lambda_i}\rho_{D,\lambda_j'\lambda_j}
\quad\mbox{(sum convention used)}\label{B_2}
\end{equation}
 is thus composed of the
(unnormalized) spin density production matrix
\begin{equation}
 \rho_P^{\lambda_i\lambda_j \lambda_i'\lambda_j'}=
T_P^{\lambda_i \lambda_j} T_P^{\lambda_i'\lambda_j' *}\label{B_3}
\end{equation}
of $\tilde{\chi}^{0}_{i,j}$ and
 the decay matrices
\begin{equation}
\rho_{D,\lambda_i'\lambda_i}=
T_{D,\lambda_i} T_{D,\lambda_i'}^{*} \quad\mbox{ and}\quad  
\rho_{D,\lambda_j'\lambda_j}=
T_{D,\lambda_j} T_{D,\lambda_j'}^{*}.\label{B_4}
\end{equation} 
Introducing a suitable set of polarization vectors for each of the
neutralinos one can expand the spin density matrix of the production
process and the decay matrices of both neutralinos in terms of Pauli
matrices.

\noindent The spin density production matrix reads:
\begin{eqnarray}
\rho_P^{\lambda_i \lambda_j \lambda'_i \lambda'_j}&=&
\Big(\delta_{\lambda_i \lambda'_i}
\delta_{\lambda_j \lambda'_j} P 
+ \delta_{\lambda_j \lambda'_j}\sum_{a}\sigma^{a}_{\lambda_i \lambda'_i}
\Sigma_P^{a}(\tilde{\chi}^0_i)
+ \delta_{\lambda_i\lambda'_i}\sum_{b}\sigma^{b}_{\lambda_j\lambda'_j}
\Sigma_P^{b}(\tilde{\chi}^0_j)\nonumber\\
& &
+ \sum_{ab}\sigma^{a}_{\lambda_i\lambda'_i}\sigma^{b}_{\lambda_j\lambda'_j}
\Sigma_P^{ab}(\tilde{\chi}^0_i\tilde{\chi}^0_j)\Big),\label{B_5}
\end{eqnarray}
and the matrices for the decays read:
\begin{eqnarray}
\rho_{D,\lambda'_i\lambda_i}&=& 
\Big(\delta_{\lambda'_i\lambda_i}D(\tilde{\chi}^0_i)+
\sum_a
\sigma^a_{\lambda'_i\lambda_i}\Sigma^a_D(\tilde{\chi}^0_i)\Big),\label{B_6}\\
\rho_{D,\lambda'_j\lambda_j}&=& 
\Big(\delta_{\lambda'_j\lambda_j}D(\tilde{\chi}^0_j)+
\sum_b\sigma^b_{\lambda'_j\lambda_j}\Sigma^b_D(\tilde{\chi}^0_j)\Big).
\label{B_7}
\end{eqnarray}
\end{appendix}

\newpage

\begin{figure}
\begin{minipage}[t]{6cm}
\begin{center}
{\setlength{\unitlength}{1cm}
\begin{picture}(5,2.5)
\put(-3.8,-7.3){\includegraphics{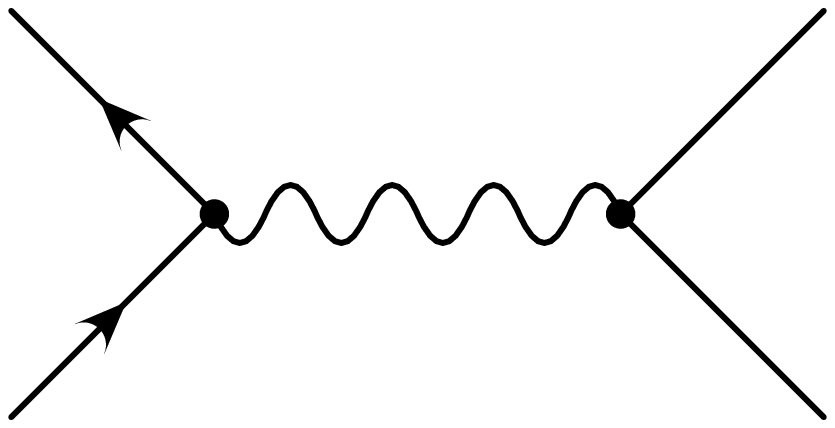}}
\put(-1,-.2){$e^{-}(p_1)$}
\put(5.2,-.2){$\tilde{\chi}^0_j(p_4)$}
\put(-1,2.3){$e^{+}(p_2)$}
\put(5.2,2.3){$\tilde{\chi}^0_i(p_3)$}
\put(2.5,.5){$Z^0$}
\end{picture}}
\label{s-channel}
\end{center}
\end{minipage}
\hspace{2cm}
\vspace{.8cm}
\begin{minipage}[t]{5cm}
\begin{center}
{\setlength{\unitlength}{1cm}
\begin{picture}(5,2.5)
\put(-3.8,-7.3){\includegraphics{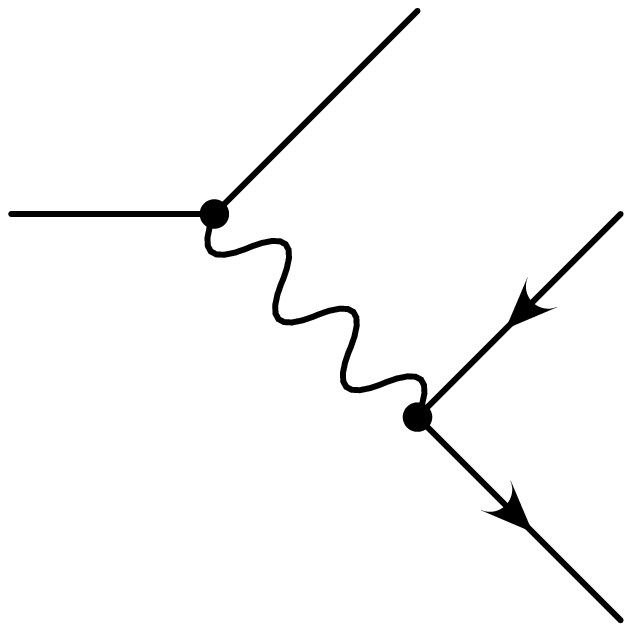}}
\put(4.8,-1.4){$\ell^{-}(p_7)$, $\ell^{-}(p_{10})$}
\put(-1,1.3){$\tilde{\chi}^0_{i}(p_3)$, $\tilde{\chi}^0_{j}(p_4)$}
\put(4.8,1.3){$\ell^{+}(p_6)$, $\ell^{+}(p_9)$}
\put(3.4,2.6){$\tilde{\chi}^0_{k}(p_5)$, $\tilde{\chi}^0_{l}(p_8)$}
\put(2,-.1){$Z^0$}
\end{picture}}
\label{sq-channel}
\end{center}
\end{minipage}
\vspace{.5cm}

\hspace{1cm}
\begin{minipage}[t]{5cm}
\begin{center}
{\setlength{\unitlength}{1cm}
\begin{picture}(2.5,5)
\put(-5.2,-6){\includegraphics{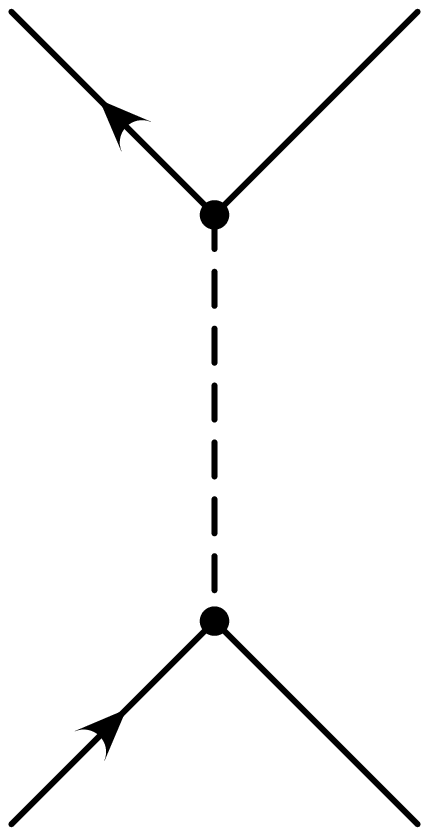}}
\put(-1.2,0){$e^{-}(p_1)$}
\put(-1.2,5){$e^{+}(p_2)$}
\put(2.7,0){$\tilde{\chi}^0_j(p_4)$}
\put(2.7,5){$\tilde{\chi}^0_i(p_3)$}
\put(1.5,2.5){$\tilde{e}_{L,R}$}
 \end{picture}}
\label{t-channel}
\end{center}
\end{minipage}
\hspace{4cm}
\begin{minipage}[t]{5cm}
\begin{center}
{\setlength{\unitlength}{1cm}
\begin{picture}(5,4)
\put(-5.2,-6){\includegraphics{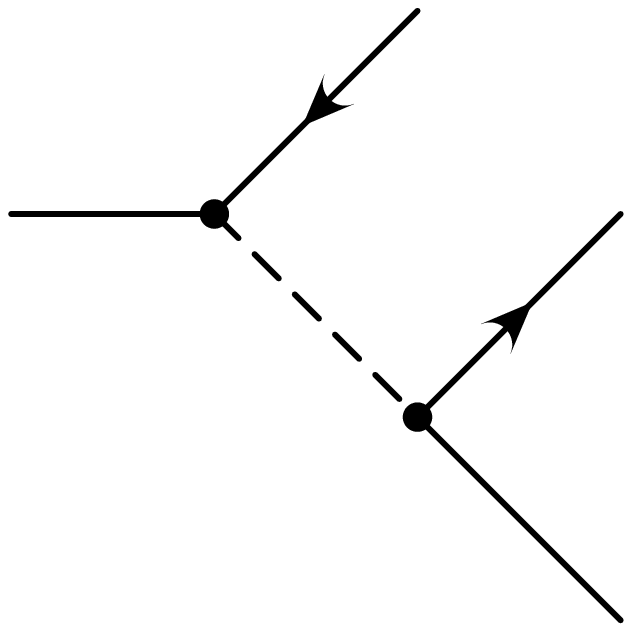}}
\put(3.5,3.1){$\ell^{-}(p_7)$, $\ell^{-}(p_{10})$}
\put(2.1,4.5){$\ell^{+}(p_6)$, $\ell^{+}(p_9)$}
\put(3.5,0.5){$\tilde{\chi}^0_{k}(p_5)$, $\tilde{\chi}^0_{l}(p_8)$}
\put(-2.3,3.3){$\tilde{\chi}^0_{i}(p_3)$, $\tilde{\chi}^0_{j}(p_4)$}
\put(.5,1.5){$\tilde{\ell}_{L,R}$}
\end{picture}}
\label{tq-channel}
\end{center}
\end{minipage}

\vspace{.5cm}
\hspace{1cm}
\begin{minipage}[t]{5cm}
\begin{center}
{\setlength{\unitlength}{1cm}
\begin{picture}(2.5,5)
\put(-5.2,-6){\includegraphics{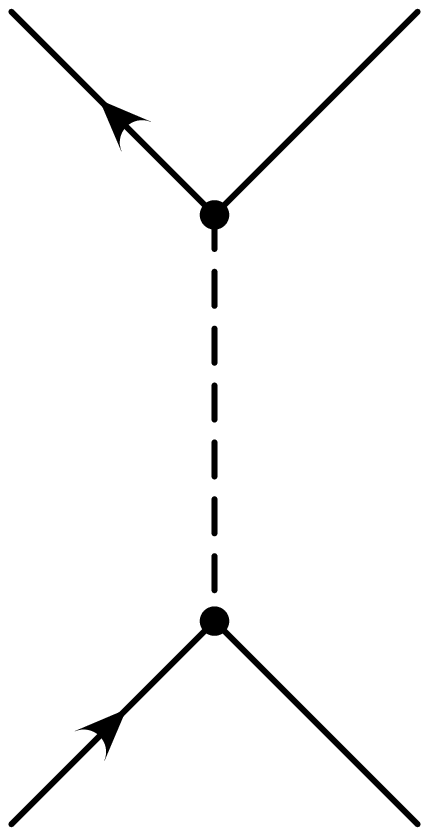}}
\put(-1.2,0){$e^{-}(p_1)$}
\put(-1.2,5){$e^{+}(p_2)$}
\put(2.7,0){$\tilde{\chi}^0_i (p_3)$}
\put(2.7,5){$\tilde{\chi}^0_j (p_4)$}
\put(1.5,2.5){$\tilde{e}_{L,R}$}
\end{picture}}
\label{u-channel}
\end{center}
\end{minipage}
\hspace{4cm}
\begin{minipage}[t]{5cm}
\begin{center}
{\setlength{\unitlength}{1cm}
\begin{picture}(5,4.5)
\put(-5.2,-6){\includegraphics{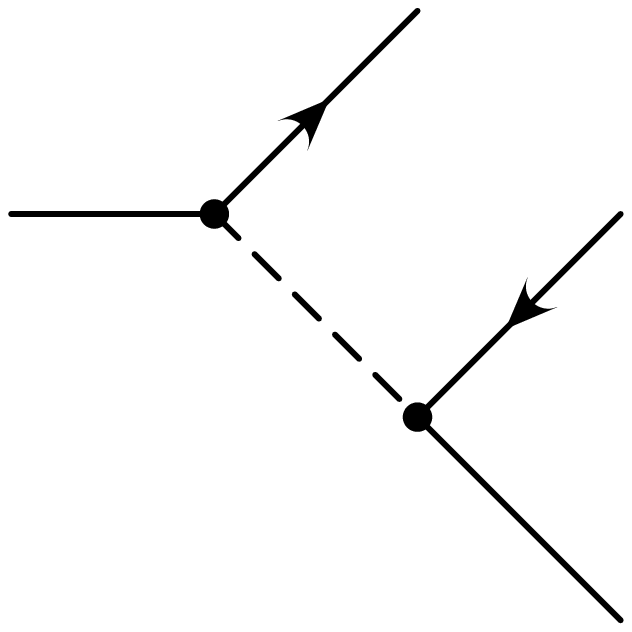}}
\put(2.1,4.5){$\ell^{-}(p_7)$, $\ell^{-}(p_{10})$}
\put(3.5,3.1){$\ell^{+}(p_6)$, $\ell^{+}(p_9)$}
\put(-2.3,3.3){$\tilde{\chi}^0_{i}(p_3)$, $\tilde{\chi}^0_{j}(p_4)$}
\put(3.5,.5){$\tilde{\chi}^0_{k}(p_5)$, $\tilde{\chi}^0_{l}(p_8)$}
\put(.5,2){$\tilde{\ell}_{L,R}$}
\end{picture}}
\label{uq-channel}
\end{center}
\end{minipage}
\caption{Feynman diagrams for the $s,t,u$ channel of the production process,
  $e^{+}e^{-}\to\tilde{\chi}^0_i\tilde{\chi}^0_j$, and the 
$s_i, t_i,u_i$ and $s_j,t_j,u_j$ channels of the decay processes
$\tilde{\chi}^0_i\to \tilde{\chi}^0_k\ell^{+}\ell^{-}$ and
$\tilde{\chi}^0_j\to \tilde{\chi}^0_l\ell^{+}\ell^{-}$.}
\end{figure}
\newpage

\begin{figure}
\begin{picture}(8,7)
\put(0,0){\includegraphics{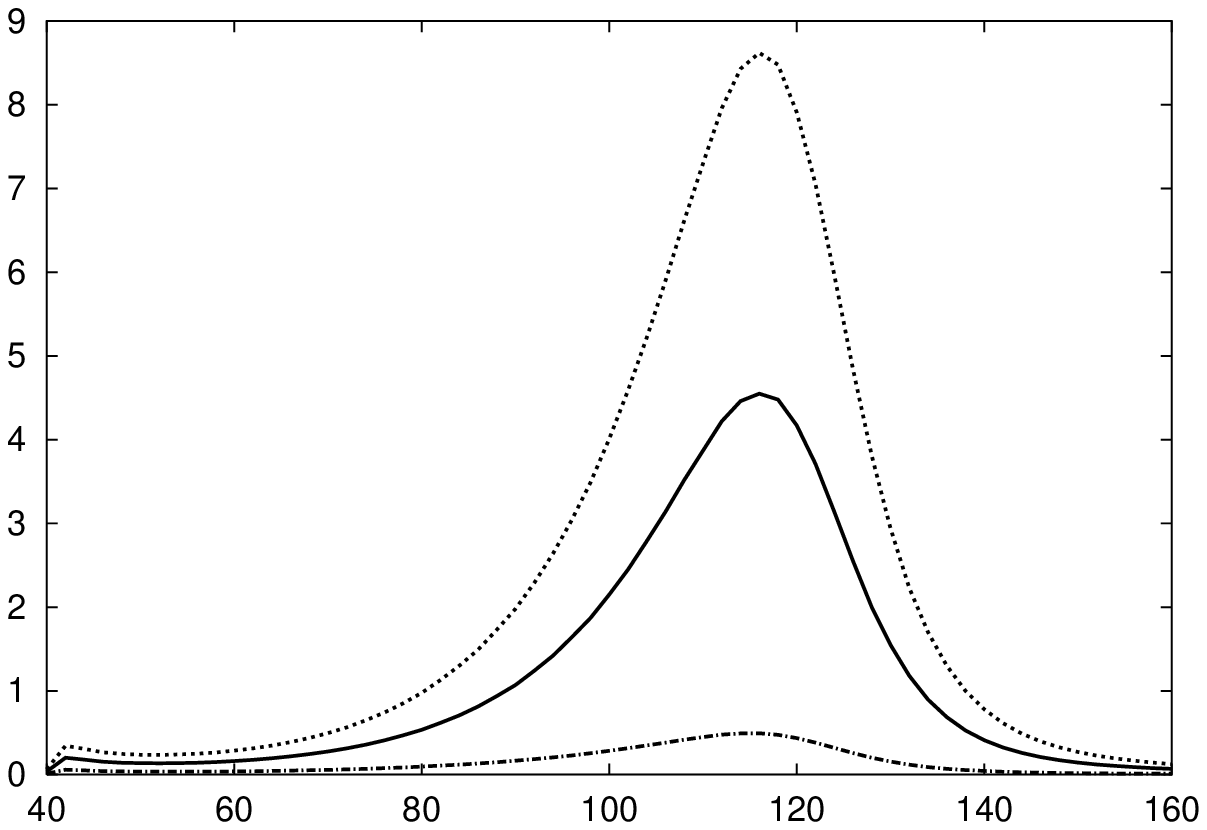}}
\put(12,1.5){$M'$~/GeV}
\put(.6,9.1){$\sigma_e/fb$}
\end{picture}
\vspace*{-1.5cm}
\caption{
$M'$ dependence of the cross section $\sigma_e$ near threshold
  ($\sqrt{s}=m_{\tilde{\chi}^0_1}+m_{\tilde{\chi}^0_2}+50$~GeV) with
$M=152$~GeV, $\mu=316$~GeV, $\tan\beta=3$, 
$m_{\tilde{e}_L}=1000$~GeV, and $m_{\tilde{e}_R}=200$~GeV for the
three cases:
unpolarized beams (solid line), $e^{-}$ beam polarized, 
$P^3_{-}=+90\%$ (dotted line) and 
$P^3_{-}=-90\%$ (dash-dotted line).
\label{sitt50_12}}
\end{figure}

\begin{figure}
\begin{picture}(8,7)
\put(0,0){\includegraphics{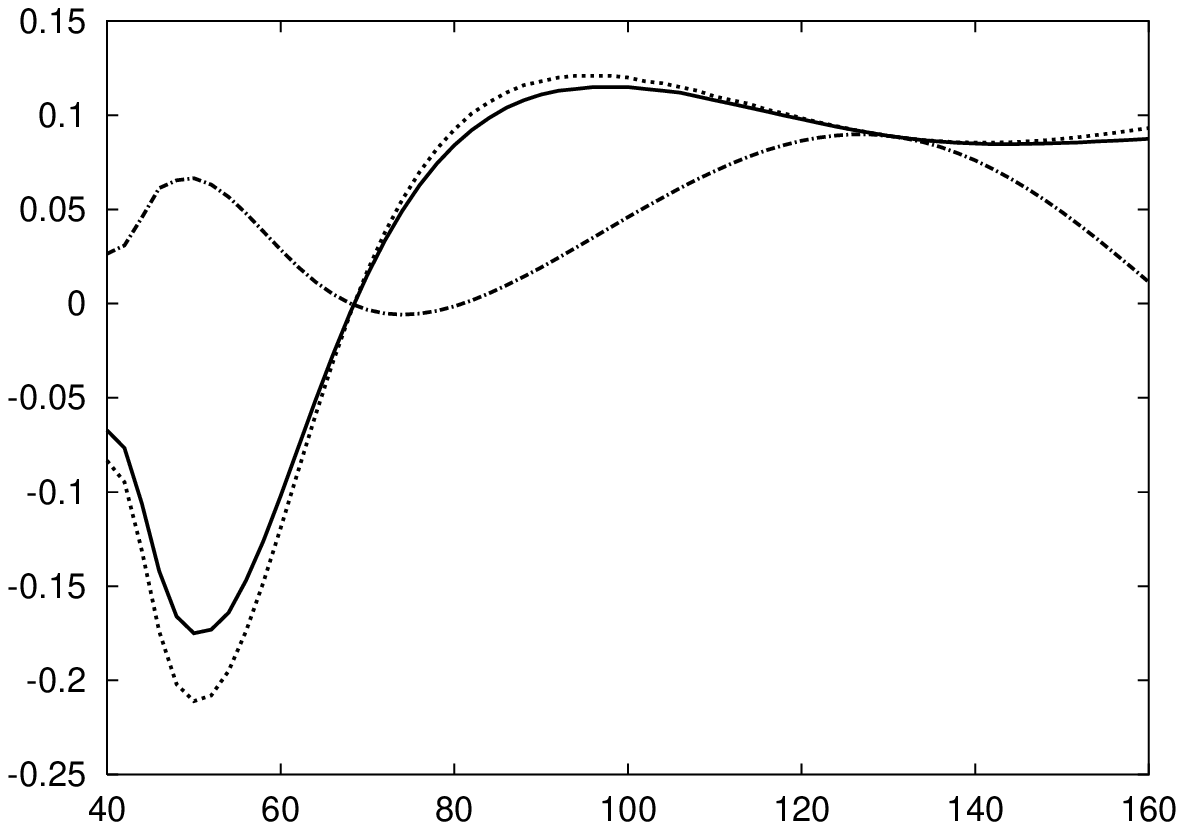}}
\put(12,1.5){$M'$~/GeV}
\put(1,9.1){$ A_{FB}$}
\end{picture}
\vspace*{-1.5cm}
\caption{
$M'$ dependence of the lepton  
forward--backward asymmetry $A_{FB}$ near threshold 
  ($\sqrt{s}=m_{\tilde{\chi}^0_1}+m_{\tilde{\chi}^0_2}+50$~GeV) with
$M=152$~GeV, $\mu=316$~GeV, $\tan\beta=3$, 
$m_{\tilde{e}_L}=1000$~GeV, and $m_{\tilde{e}_R}=200$~GeV for the
three cases:
unpolarized beams (solid line),
$e^{-}$ beam polarized, $P^3_{-}=+90\%$ (dotted line) and 
$P^3_{-}=-90\%$ (dash-dotted line).
\label{asyt50_12}}
\end{figure}

\begin{figure}
\begin{picture}(8,7)
\put(0,0){\includegraphics{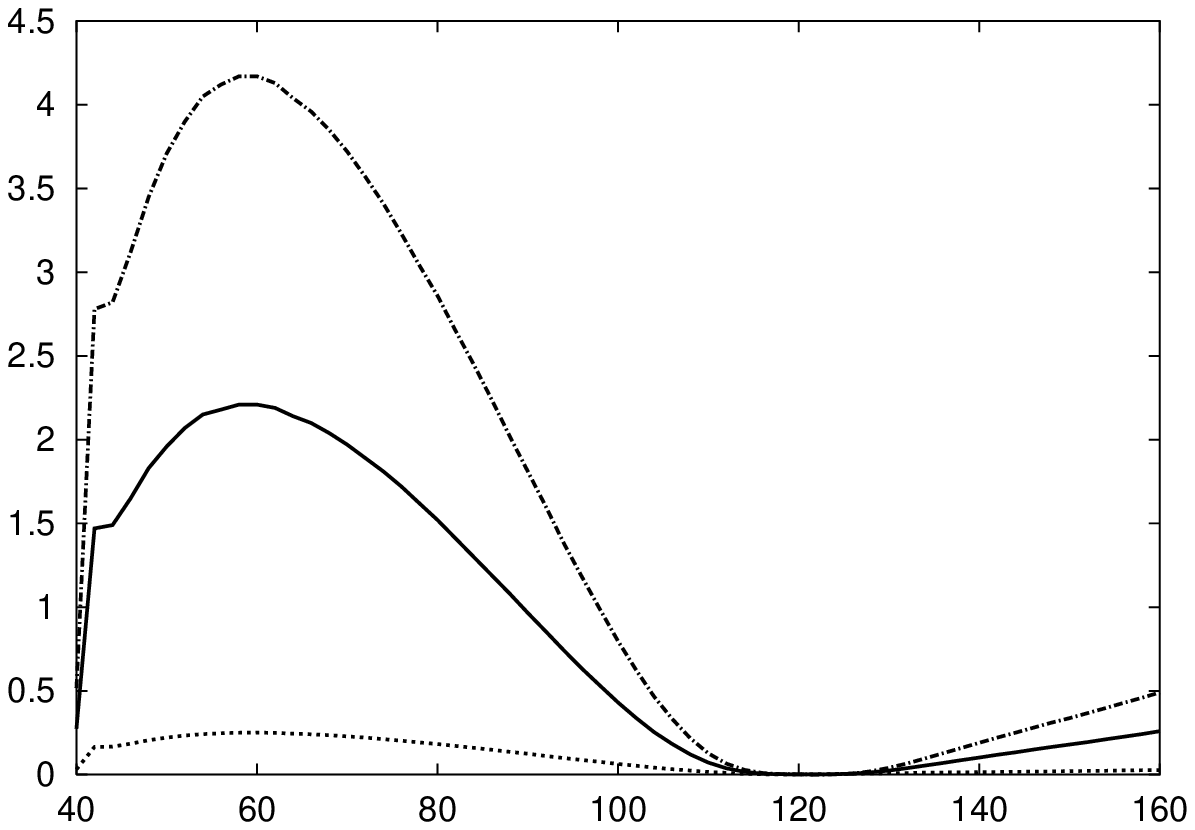}}
\put(12,1.5){$M'$}
\put(.6,9.1){$\sigma_e/fb$}
\end{picture}
\vspace*{-1.5cm}
\caption{
$M'$ dependence of the cross section $\sigma_e$ near threshold
  ($\sqrt{s}=m_{\tilde{\chi}^0_1}+m_{\tilde{\chi}^0_2}+50$~GeV) with
$M=152$~GeV, $\mu=316$~GeV, $\tan\beta=3$, 
$m_{\tilde{e}_L}=200$~GeV, and $m_{\tilde{e}_R}=1000$~GeV for the
three cases:
unpolarized beams (solid line), 
$e^{-}$ beam polarized, $P^3_{-}=+90\%$ (dotted line) and 
$P^3_{-}=-90\%$ (dash-dotted line).
\label{sitt50_21}}
\end{figure}

\begin{figure}
\begin{picture}(8,7)
\put(0,0){\includegraphics{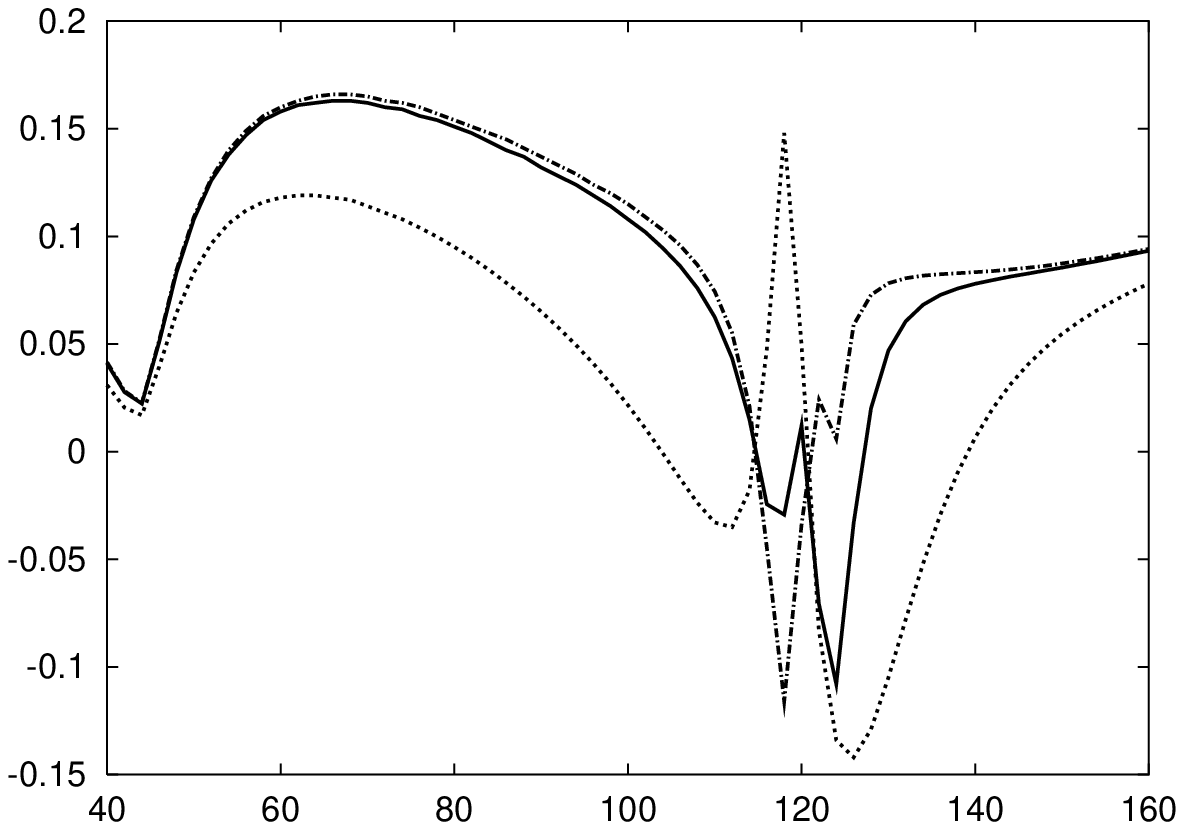}}
\put(12,1.5){$M'$}
\put(1,9.1){$ A_{FB}$}
\end{picture}
\vspace*{-1.5cm}
\caption{
$M'$ dependence of the lepton  
forward--backward asymmetry $A_{FB}$ near threshold 
  ($\sqrt{s}=m_{\tilde{\chi}^0_1}+m_{\tilde{\chi}^0_2}+50$~GeV) with
$M=152$~GeV, $\mu=316$~GeV, $\tan\beta=3$, 
$m_{\tilde{e}_L}=200$~GeV, and $m_{\tilde{e}_R}=1000$~GeV for the
three cases:
unpolarized beams (solid line),  
$e^{-}$ beam polarized, $P^3_{-}=+90\%$ (dotted line) and 
$P^3_{-}=-90\%$ (dash-dotted line).
\label{asyt50_21}}
\end{figure}

\begin{figure}
\begin{picture}(8,7)
\put(0,0){\includegraphics{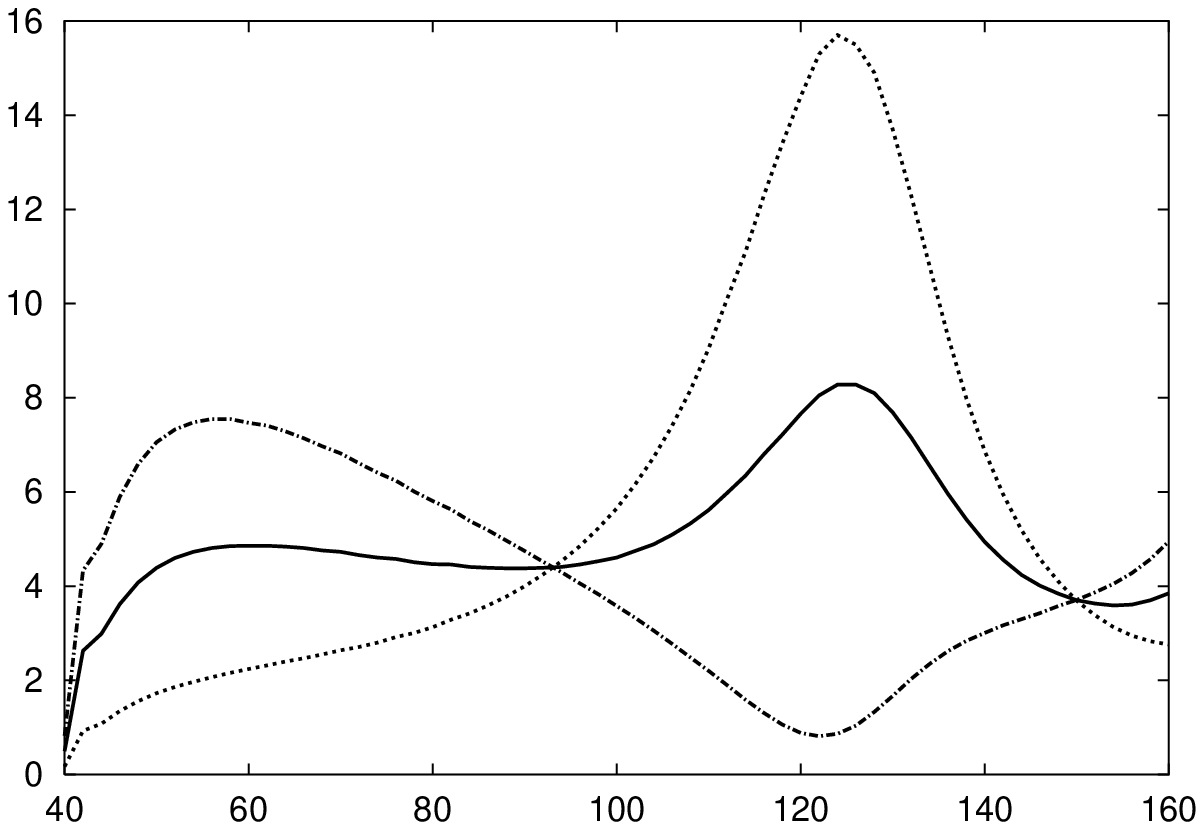}}
\put(12,1.5){$ M'$~/GeV}
\put(.6,9.1){$\sigma_e/fb$}
\end{picture}
\vspace*{-1.5cm}
\caption{
$M'$ dependence of the cross section $\sigma_e$ near threshold
  ($\sqrt{s}=m_{\tilde{\chi}^0_1}+m_{\tilde{\chi}^0_2}+50$~GeV) with
$M=152$~GeV, $\mu=316$~GeV, $\tan\beta=3$, 
$m_{\tilde{e}_L}=176$~GeV, and $m_{\tilde{e}_R}=161$~GeV for the
three cases:
unpolarized beams (solid line), 
$e^{-}$ beam polarized, $P^3_{-}=+90\%$ (dotted line) and 
$P^3_{-}=-90\%$ (dash-dotted line).
\label{sitt50_m0}}
\end{figure}

\begin{figure}
\begin{picture}(8,7)
\put(0,0){\includegraphics{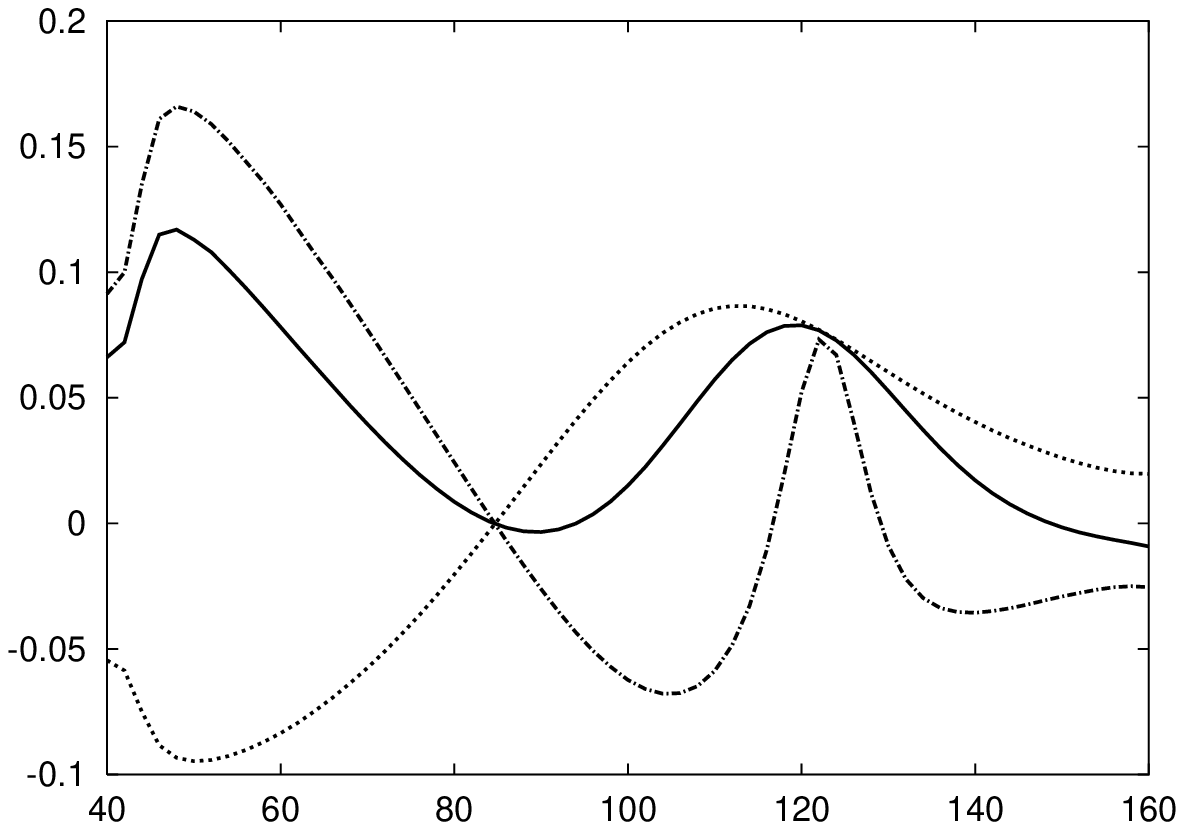}}
\put(12,1.5){$M'$~/GeV}
\put(1,9.1){$A_{FB}$}
\end{picture}
\vspace*{-1.5cm}
\caption{
$M'$ dependence of the lepton  
forward--backward asymmetry $A_{FB}$ near threshold
($\sqrt{s}=m_{\tilde{\chi}^0_1}+m_{\tilde{\chi}^0_2}+50$~GeV) with
$M=152$~GeV, $\mu=316$~GeV, $\tan\beta=3$, 
$m_{\tilde{e}_L}=176$~GeV, and $m_{\tilde{e}_R}=161$~GeV for the
three cases:
unpolarized beams (solid line), 
$e^{-}$ beam polarized, $P^3_{-}=+90\%$ (dotted line) and 
$P^3_{-}=-90\%$ (dash-dotted line).
\label{asyt50_m0}}
\end{figure}

\begin{figure}
\begin{picture}(8,7)
\put(0,0){\includegraphics{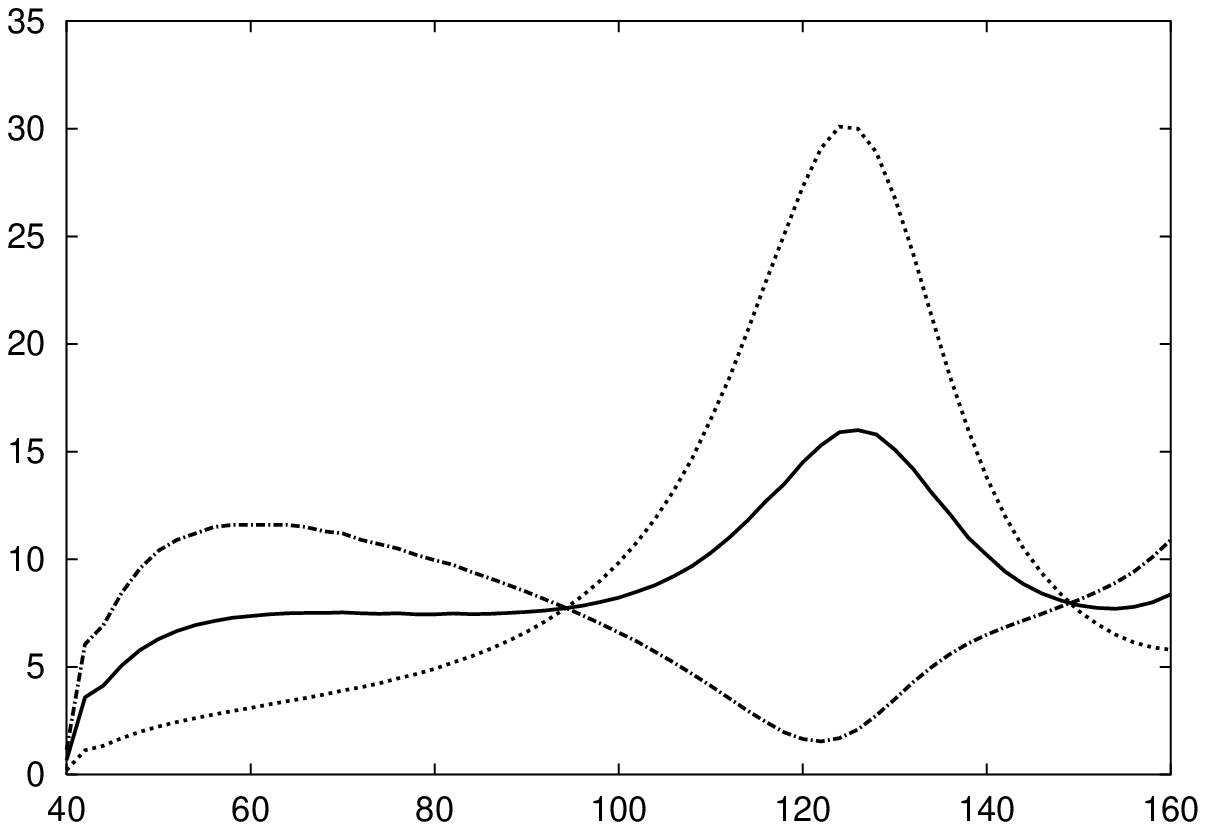}}
\put(12,1.5){$ M' /GeV$}
\put(.6,9.1){$\sigma_e/fb$}
\end{picture}
\vspace*{-1.5cm}
\caption{$M'$ dependence of the cross section $\sigma_e$ at
  $\sqrt{s}=500$~GeV with
$M=152$~GeV, $\mu=316$~GeV, $\tan\beta=3$, 
$m_{\tilde{e}_L}=176$~GeV, and $m_{\tilde{e}_R}=161$~GeV for the
three cases:
unpolarized beams (solid line), 
$e^{-}$ beam polarized, $P^3_{-}=+90\%$ (dotted line) and 
$P^3_{-}=-90\%$ (dash-dotted line).
\label{sit_m0}}
\end{figure}

\begin{figure}
\begin{picture}(8,7)
\put(0,0){\includegraphics{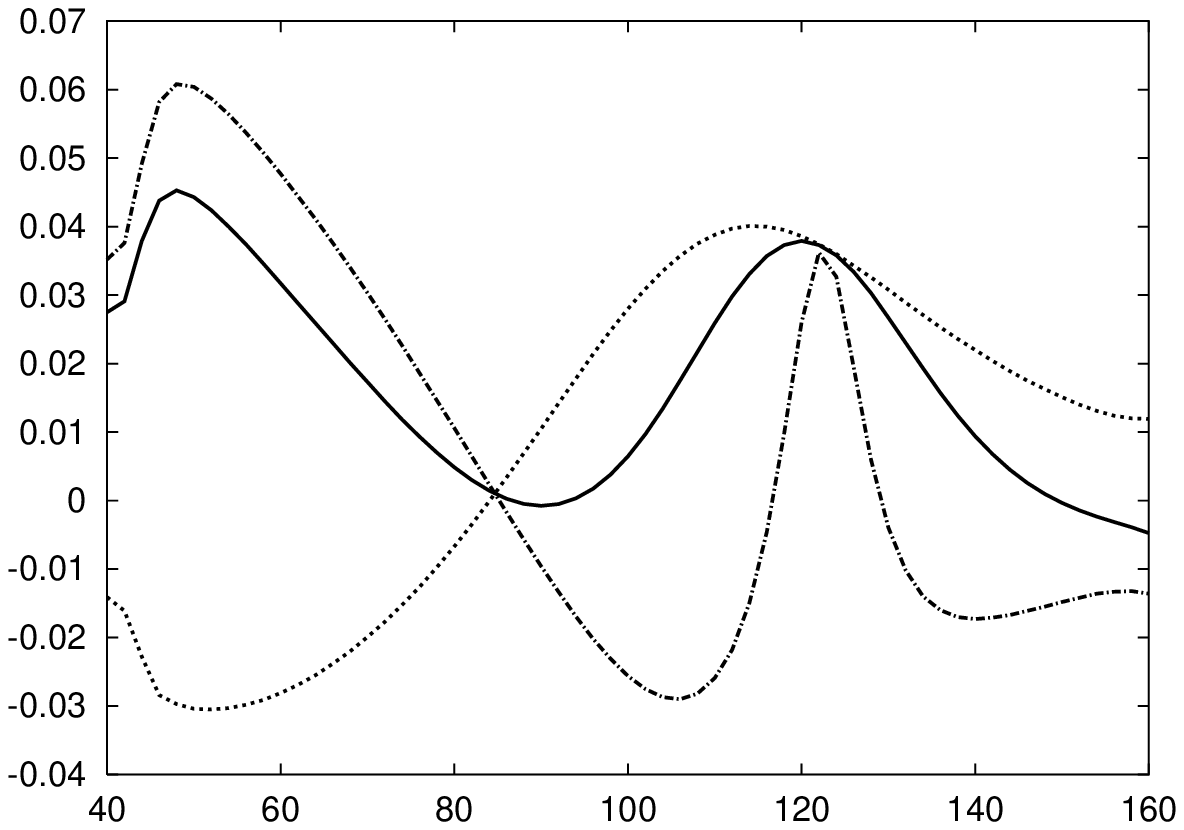}}
\put(12,1.5){$M'$~/GeV}
\put(1,9.1){$ A_{FB}$}
\end{picture}
\vspace*{-1.5cm}
\caption{$M'$ dependence of the lepton 
forward--backward asymmetry $A_{FB}$ at
  $\sqrt{s}=500$~GeV  with
$M=152$~GeV, $\mu=316$~GeV, $\tan\beta=3$, 
$m_{\tilde{e}_L}=176$~GeV, and $m_{\tilde{e}_R}=161$~GeV for the
three cases:
unpolarized beams (solid line),
$e^{-}$ beam polarized, $P^3_{-}=+90\%$ (dotted line) and 
$P^3_{-}=-90\%$ (dash-dotted line).
\label{asy_m0}}
\end{figure}

\end{document}